%% file: paper.tex
\documentclass[sigconf]{acmart}

\AtBeginDocument{%
  \providecommand\BibTeX{{%
    \normalfont B\kern-0.5em{\scshape i\kern-0.25em b}\kern-0.8em\TeX}}}

\copyrightyear{2022}
\acmYear{2022}
\setcopyright{acmcopyright}
\acmConference[WiPSCE '22]{Proceedings of the 17th Workshop in Primary and Secondary Computing Education}{October 31-November 2, 2022}{Morschach, Switzerland}
\acmBooktitle{Proceedings of the 17th Workshop in Primary and Secondary Computing Education (WiPSCE '22), October 31-November 2, 2022, Morschach, Switzerland}
\acmPrice{15.00}
\acmDOI{10.1145/3556787.3556860}
\acmISBN{978-1-4503-9853-4/22/10}

\newcommand{\summary}[2]{
        \vspace{0.4em}
        \noindent
        \colorbox{gray!20}{%
            \parbox{.97\linewidth}{%
                    \textbf{\textsf{Summary (\textit{#1})}}
                #2
            }%
        }%
}%

\input{packages}

\begin{document}

\input{macros}

\input{generated_files/stats}

 \title[Programming Ozobots in Primary School]{
Common Problems and Effects of Feedback on Fun \\ When Programming Ozobots in Primary School}


\author{Luisa Greifenstein}
\affiliation{%
	\institution{University of Passau}
	\city{Passau}%
	\country{Germany}
}

\author{Isabella Graßl}
\affiliation{%
	\institution{University of Passau}
	\city{Passau}%
	\country{Germany}
}

\author{Ute Heuer}
\affiliation{%
	\institution{University of Passau}
	\city{Passau}%
	\country{Germany}
}

\author{Gordon Fraser}
\affiliation{%
	\institution{University of Passau}
	\city{Passau}
	\country{Germany}
}

\renewcommand{\shortauthors}{Greifenstein et al.}

\begin{abstract}
\looseness=-1
Computational thinking is increasingly introduced at primary school level, usually with some form of programming activity. In particular, educational robots provide an opportunity for engaging students with programming through hands-on experiences. 
However, primary school teachers might not be adequately prepared for teaching computer science related topics, 
and giving feedback to students can often be challenging: Besides the content of the feedback (e.g., what problems have to be handled), the way the feedback is given is also important, as it can lead to negative emotional effects.
To support teachers with the way of giving feedback on common problems when teaching programming with robotics, we conducted a study consisting of \numWorkshops workshops with \numThirdClasses third and \numFourthClasses fourth grade primary school classes. Within seven different activities, the \numTotalParticipants primary school children first programmed the Ozobot Evo robot in the pen-and-paper mode and then on a digital device. Throughout these activities we collected data on the problems the students encountered, the feedback given, and the fun they experienced.
Our analysis reveals eight categories of problems, which we summarise in this paper together with corresponding possible feedback. We observed that problems that are urgent or can harm the students' self-efficacy have a negative impact on how enjoyable an activity is perceived. While direct instruction significantly decreased the experienced fun, hints had a positive effect. Generally, we found programming the Ozobot Evo to be encouraging for both girls and boys. 
To support teachers, we discuss ideas for giving encouraging feedback on common problems of Ozobot Evo programming activities and how our findings transfer to other robots.
\end{abstract}




\maketitle

\input{content/introduction}

\input{content/relatedwork}
\input{content/workshop}

\input{content/method}

\input{content/results}

\input{content/discussion}
\input{content/conclusions}

\vspace{-0.3em}
\begin{acks}
\vspace{-0.3em}This work is supported by the Federal Ministry of Education and Research
through project ``primary::programming'' (01JA2021) as
part of the ``Qualitätsoffensive Lehrerbildung'', a joint initiative of the
Federal Government and the Länder. The authors are responsible for the content
of this publication.
\end{acks}

\balance

\bibliographystyle{ACM-Reference-Format}
\bibliography{references}

\end{document}

%% file: packages.tex
\usepackage{xspace}

\usepackage[capitalise]{cleveref}
\usepackage{csquotes}
\usepackage{fontawesome}

\usepackage{tabularx}
\usepackage{colortbl}
\usepackage{subfig}
\usepackage{multirow}
\usepackage{listings}
\usepackage{xcolor}
\usepackage{fp}

\usepackage{collcell}

\colorlet{punct}{red!60!black}
\definecolor{background}{HTML}{EEEEEE}
\definecolor{delim}{RGB}{20,105,176}
\colorlet{numb}{magenta!60!black}

\definecolor{Gray}{gray}{0.9}

\lstdefinelanguage{json}{
    basicstyle=\normalfont\ttfamily,
    numbers=left,
    numberstyle=\scriptsize,
    stepnumber=1,
    numbersep=8pt,
    showstringspaces=false,
    breaklines=true,
    frame=lines,
    backgroundcolor=\color{background},
    literate=
     *{0}{{{\color{numb}0}}}{1}
      {1}{{{\color{numb}1}}}{1}
      {2}{{{\color{numb}2}}}{1}
      {3}{{{\color{numb}3}}}{1}
      {4}{{{\color{numb}4}}}{1}
      {5}{{{\color{numb}5}}}{1}
      {6}{{{\color{numb}6}}}{1}
      {7}{{{\color{numb}7}}}{1}
      {8}{{{\color{numb}8}}}{1}
      {9}{{{\color{numb}9}}}{1}
      {:}{{{\color{punct}{:}}}}{1}
      {,}{{{\color{punct}{,}}}}{1}
      {\{}{{{\color{delim}{\{}}}}{1}
      {\}}{{{\color{delim}{\}}}}}{1}
      {[}{{{\color{delim}{[}}}}{1}
      {]}{{{\color{delim}{]}}}}{1},
}

\DeclareQuoteStyle[american]{english}
        {\itshape\textquotedblleft}
        [\textquotedblleft]
        {\textquotedblright}
        [0.05em]
        {\textquoteleft}
        {\textquoteright}


%% file: macros.tex
\newcommand{\numWorkshops}{seven\xspace}
\newcommand{\numThirdClasses}{three\xspace}
\newcommand{\numFourthClasses}{four\xspace}
\newcommand{\wernstein}{\textit{blinded}\xspace}

\newcommand{\irr}{0.65\xspace}

\newcommand{\groupConstellations}{\numTwoFemalesGroups all-female groups, \numTwoMalesGroups all-male groups, \numFemaleMaleGroups mixed groups}

\newcommand{\rqOne}{Where do primary school children need support with Ozobot Evo programming activities?
\xspace}

\newcommand{\rqTwo}{Do primary school children that receive feedback on a specific problem or via a specific tutoring component have less fun?\xspace
}

\newcommand{\rqThree}{Does gender have an effect on problems or fun?\xspace
}


\newcommand\zz[1]{%
\ifdim#1pt<5pt\cellcolor{white}\else
\ifdim#1pt<10pt\cellcolor{gray!10}\else
\ifdim#1pt<15pt\cellcolor{gray!20}\else
\ifdim#1pt<20pt\cellcolor{gray!30}\else
\ifdim#1pt<25pt\cellcolor{gray!40}\else
\ifdim#1pt<30pt\cellcolor{gray!50}\else
\ifdim#1pt<35pt\cellcolor{gray!60}\else
\ifdim#1pt<40pt\cellcolor{gray!70}\else
\ifdim#1pt<45pt\cellcolor{gray}\else
\cellcolor{white}\fi\fi\fi\fi\fi\fi\fi\fi\fi
#1}
\newcolumntype{C}{>{\collectcell\zz}c<{\endcollectcell}}

\newcommand{\delP}[1]{\StrSubstitute{#1}{p}{}}
\newcommand{\delG}[1]{\StrSubstitute{#1}{p=}{}}

%% file: generated_files/stats.tex
\newcommand{\numTwoFemalesGroups}{19\xspace}
\newcommand{\numTwoMalesGroups}{23\xspace}
\newcommand{\numFemaleMaleGroups}{12\xspace}
\newcommand{\numGroups}{54\xspace}
\newcommand{\numOneFemaleGroups}{5\xspace}
\newcommand{\numOneMaleGroups}{3\xspace}
\newcommand{\numOneStudentGroups}{8\xspace}
\newcommand{\numTotalParticipants}{116\xspace}
\newcommand{\numGirls}{55\xspace}
\newcommand{\numBoys}{61\xspace}
\newcommand{\numParticipantsThirdGrade}{41\xspace}
\newcommand{\numParticipantsFourthGrade}{75\xspace}
\newcommand{\againOneagainFivepValue}{0.023\xspace}
\newcommand{\againOneagainFiveEffectSize}{0.18\xspace}
\newcommand{\againThreeagainFivepValue}{0.025\xspace}
\newcommand{\againThreeagainFiveEffectSize}{0.18\xspace}
\newcommand{\againOneagainSevenpValue}{0.02\xspace}
\newcommand{\againOneagainSevenEffectSize}{0.19\xspace}
\newcommand{\againThreeagainSevenpValue}{0.021\xspace}
\newcommand{\againThreeagainSevenEffectSize}{0.19\xspace}
\newcommand{\againOneagainSixpValue}{0.121\xspace}
\newcommand{\againOneagainSixEffectSize}{0.11\xspace}
\newcommand{\againThreeagainSixpValue}{0.125\xspace}
\newcommand{\againThreeagainSixEffectSize}{0.11\xspace}
\newcommand{\againFouragainFivepValue}{0.083\xspace}
\newcommand{\againFouragainFiveEffectSize}{0.13\xspace}
\newcommand{\againFouragainSevenpValue}{0.071\xspace}
\newcommand{\againFouragainSevenEffectSize}{0.14\xspace}
\newcommand{\likeFourlikeSevenpValue}{0.198\xspace}
\newcommand{\likeFourlikeSevenEffectSize}{0.08\xspace}
\newcommand{\likeFourlikeSixpValue}{0.467\xspace}
\newcommand{\likeFourlikeSixEffectSize}{0.01\xspace}
\newcommand{\likeThreelikeSixpValue}{0.382\xspace}
\newcommand{\likeThreelikeSixEffectSize}{0.03\xspace}
\newcommand{\againOneGenderpValue}{0.485\xspace}
\newcommand{\againOneGenderEffectSize}{0\xspace}
\newcommand{\againTwoGenderpValue}{0.61\xspace}
\newcommand{\againTwoGenderEffectSize}{0.03\xspace}
\newcommand{\againThreeGenderpValue}{0.472\xspace}
\newcommand{\againThreeGenderEffectSize}{0.01\xspace}
\newcommand{\againFourGenderpValue}{0.184\xspace}
\newcommand{\againFourGenderEffectSize}{0.08\xspace}
\newcommand{\againFiveGenderpValue}{0.019\xspace}
\newcommand{\againFiveGenderEffectSize}{0.19\xspace}
\newcommand{\againSixGenderpValue}{0.335\xspace}
\newcommand{\againSixGenderEffectSize}{0.04\xspace}
\newcommand{\againSevenGenderpValue}{0.792\xspace}
\newcommand{\againSevenGenderEffectSize}{0.08\xspace}
\newcommand{\likeOneGenderpValue}{0.855\xspace}
\newcommand{\likeOneGenderEffectSize}{0.1\xspace}
\newcommand{\likeTwoGenderpValue}{0.402\xspace}
\newcommand{\likeTwoGenderEffectSize}{0.02\xspace}
\newcommand{\likeThreeGenderpValue}{0.965\xspace}
\newcommand{\likeThreeGenderEffectSize}{0.17\xspace}
\newcommand{\likeFourGenderpValue}{0.176\xspace}
\newcommand{\likeFourGenderEffectSize}{0.09\xspace}
\newcommand{\likeFiveGenderpValue}{0.077\xspace}
\newcommand{\likeFiveGenderEffectSize}{0.13\xspace}
\newcommand{\likeSixGenderpValue}{0.168\xspace}
\newcommand{\likeSixGenderEffectSize}{0.09\xspace}
\newcommand{\likeSevenGenderpValue}{0.122\xspace}
\newcommand{\likeSevenGenderEffectSize}{0.11\xspace}
\newcommand{\againOneGroupspValue}{0.398\xspace}
\newcommand{\againOneGroupsEffectSize}{0.04\xspace}
\newcommand{\againTwoGroupspValue}{0.318\xspace}
\newcommand{\againThreeGroupspValue}{0.937\xspace}
\newcommand{\againFourGroupspValue}{0.439\xspace}
\newcommand{\againFiveroupspValue}{0.219\xspace}
\newcommand{\againSixGroupspValue}{0.57\xspace}
\newcommand{\againSevenGroupspValue}{0.44\xspace}
\newcommand{\likeOneGroupspValue}{0.966\xspace}
\newcommand{\likeTwoGroupspValue}{0.705\xspace}
\newcommand{\likeThreeGroupspValue}{0.829\xspace}
\newcommand{\likeFourGroupspValue}{0.157\xspace}
\newcommand{\likeFiveGroupspValue}{0.5\xspace}
\newcommand{\likeSixGroupspValue}{0.448\xspace}
\newcommand{\likeSevenGroupspValue}{0.526\xspace}
\newcommand{\amountInteractionsGroupspValue}{0.312\xspace}
\newcommand{\amountInteractionsGroupsEffectSize}{0.01\xspace}
\newcommand{\amountInteractionsGroupsTwoMalesMean}{3.3\xspace}
\newcommand{\amountInteractionsGroupsTwoFemalesMean}{4.37\xspace}
\newcommand{\amountInteractionsGroupsMixedMean}{3.83\xspace}
\newcommand{\problemUnderstandingTaskGroupspValue}{0.154\xspace}
\newcommand{\problemUnderstandingTaskGroupsEffectSize}{0.03\xspace}
\newcommand{\problemUnderstandingTaskGroupsTwoMales}{0.39\xspace}
\newcommand{\problemUnderstandingTaskGroupsTwoFemales}{0.68\xspace}
\newcommand{\problemUnderstandingTaskGroupsMixed}{0.33\xspace}
\newcommand{\problemUnderstandingTaskGroupsOneMale}{0\xspace}
\newcommand{\problemUnderstandingTaskGroupsOneFemale}{1\xspace}
\newcommand{\problemWrongCodeGroupspValue}{0.888\xspace}
\newcommand{\problemWrongCodeGroupsEffectSize}{-0.03\xspace}
\newcommand{\problemWrongCodeGroupsTwoMales}{0.52\xspace}
\newcommand{\problemWrongCodeGroupsTwoFemales}{0.47\xspace}
\newcommand{\problemWrongCodeGroupsMixed}{0.5\xspace}
\newcommand{\problemWrongCodeGroupsOneMale}{0.67\xspace}
\newcommand{\problemWrongCodeGroupsOneFemale}{0.4\xspace}
\newcommand{\problemOzobotSpecificGroupspValue}{0.447\xspace}
\newcommand{\problemOzobotSpecificGroupsEffectSize}{-0.01\xspace}
\newcommand{\problemOzobotSpecificGroupsTwoMales}{0.26\xspace}
\newcommand{\problemOzobotSpecificGroupsTwoFemales}{0.42\xspace}
\newcommand{\problemOzobotSpecificGroupsMixed}{0.5\xspace}
\newcommand{\problemOzobotSpecificGroupsOneMale}{0.67\xspace}
\newcommand{\problemOzobotSpecificGroupsOneFemale}{0.8\xspace}
\newcommand{\problemReadingDirectionGroupspValue}{0.093\xspace}
\newcommand{\problemReadingDirectionGroupsEffectSize}{0.05\xspace}
\newcommand{\problemReadingDirectionGroupsTwoMales}{0.26\xspace}
\newcommand{\problemReadingDirectionGroupsTwoFemales}{0.58\xspace}
\newcommand{\problemReadingDirectionGroupsMixed}{0.5\xspace}
\newcommand{\problemReadingDirectionGroupsOneMale}{0\xspace}
\newcommand{\problemReadingDirectionGroupsOneFemale}{0.6\xspace}
\newcommand{\problemUTurnGroupspValue}{0.806\xspace}
\newcommand{\problemUTurnGroupsEffectSize}{-0.03\xspace}
\newcommand{\problemUTurnGroupsTwoMales}{0.39\xspace}
\newcommand{\problemUTurnGroupsTwoFemales}{0.37\xspace}
\newcommand{\problemUTurnGroupsMixed}{0.5\xspace}
\newcommand{\problemUTurnGroupsOneMale}{0.33\xspace}
\newcommand{\problemUTurnGroupsOneFemale}{0.2\xspace}
\newcommand{\problemDirectionGroupspValue}{0.029\xspace}
\newcommand{\problemDirectionGroupsEffectSize}{0.1\xspace}
\newcommand{\problemDirectionGroupsTwoMales}{0.17\xspace}
\newcommand{\problemDirectionGroupsTwoFemales}{0.32\xspace}
\newcommand{\problemDirectionGroupsMixed}{0.58\xspace}
\newcommand{\problemDirectionGroupsOneMale}{0.33\xspace}
\newcommand{\problemDirectionGroupsOneFemale}{0\xspace}
\newcommand{\problemAppGroupspValue}{0.651\xspace}
\newcommand{\problemAppGroupsEffectSize}{-0.02\xspace}
\newcommand{\problemAppGroupsTwoMales}{0.22\xspace}
\newcommand{\problemAppGroupsTwoFemales}{0.26\xspace}
\newcommand{\problemAppGroupsMixed}{0.33\xspace}
\newcommand{\problemAppGroupsOneMale}{0.33\xspace}
\newcommand{\problemAppGroupsOneFemale}{0.2\xspace}
\newcommand{\problemOtherGroupspValue}{0.341\xspace}
\newcommand{\problemOtherGroupsEffectSize}{0\xspace}
\newcommand{\problemOtherGroupsTwoMales}{0.13\xspace}
\newcommand{\problemOtherGroupsTwoFemales}{0.05\xspace}
\newcommand{\problemOtherGroupsMixed}{0\xspace}
\newcommand{\problemOtherGroupsOneMale}{0\xspace}
\newcommand{\problemOtherGroupsOneFemale}{0\xspace}
\newcommand{\feedbackFunAllActivitiespValue}{0.311\xspace}
\newcommand{\feedbackFunAllActivitiesEffectSize}{0.05\xspace}
\newcommand{\feedbackFunActivityOnepValue}{0.051\xspace}
\newcommand{\feedbackFunActivityOneEffectSize}{0.44\xspace}
\newcommand{\feedbackFunActivityThreepValue}{0.325\xspace}
\newcommand{\feedbackFunActivityThreeEffectSize}{0.12\xspace}
\newcommand{\feedbackFunActivityFourpValue}{0.712\xspace}
\newcommand{\feedbackFunActivityFourEffectSize}{0.15\xspace}
\newcommand{\feedbackFunActivityFivepValue}{0.567\xspace}
\newcommand{\feedbackFunActivityFiveEffectSize}{0.05\xspace}
\newcommand{\feedbackFunActivitySixpValue}{0.696\xspace}
\newcommand{\feedbackFunActivitySixEffectSize}{0.14\xspace}
\newcommand{\feedbackFunActivitySevenpValue}{0.983\xspace}
\newcommand{\feedbackFunActivitySevenEffectSize}{0.56\xspace}
\newcommand{\allActivitiesAgainMean}{1.9\xspace}
\newcommand{\allActivitiesLikeMean}{3.72\xspace}
\newcommand{\allActivitiesAgainFemaleMean}{1.9\xspace}
\newcommand{\allActivitiesAgainMaleMean}{1.9\xspace}
\newcommand{\allActivitiesFemaleLikeMean}{3.74\xspace}
\newcommand{\allActivitiesMaleLikeMean}{3.71\xspace}
\newcommand{\allActivitiesTwoFemalesAgainMean}{1.9\xspace}
\newcommand{\allActivitiesTwoMalesAgainMean}{1.9\xspace}
\newcommand{\allActivitiesFemaleMaleAgainMean}{1.88\xspace}
\newcommand{\allActivitiesTwoFemalesLikeMean}{3.75\xspace}
\newcommand{\allActivitiesTwoMalesLikeMean}{3.69\xspace}
\newcommand{\allActivitiesFemaleMaleLikeMean}{3.72\xspace}
\newcommand{\againOneAllMean}{1.95\xspace}
\newcommand{\againTwoAllMean}{1.91\xspace}
\newcommand{\againThreeAllMean}{1.95\xspace}
\newcommand{\againFourAllMean}{1.93\xspace}
\newcommand{\againFiveAllMean}{1.85\xspace}
\newcommand{\againSixAllMean}{1.88\xspace}
\newcommand{\againSevenAllMean}{1.85\xspace}
\newcommand{\likeOneAllMean}{3.78\xspace}
\newcommand{\likeTwoAllMean}{3.78\xspace}
\newcommand{\likeThreeAllMean}{3.72\xspace}
\newcommand{\likeFourAllMean}{3.81\xspace}
\newcommand{\likeFiveAllMean}{3.7\xspace}
\newcommand{\likeSixAllMean}{3.69\xspace}
\newcommand{\likeSevenAllMean}{3.6\xspace}
\newcommand{\againOneMaleMean}{1.93\xspace}
\newcommand{\againTwoMaleMean}{1.9\xspace}
\newcommand{\againThreeMaleMean}{1.93\xspace}
\newcommand{\againFourMaleMean}{1.9\xspace}
\newcommand{\againFiveMaleMean}{1.93\xspace}
\newcommand{\againSixMaleMean}{1.85\xspace}
\newcommand{\againSevenMaleMean}{1.86\xspace}
\newcommand{\againOneFemaleMean}{1.96\xspace}
\newcommand{\againTwoFemaleMean}{1.93\xspace}
\newcommand{\againThreeFemaleMean}{1.96\xspace}
\newcommand{\againFourFemaleMean}{1.96\xspace}
\newcommand{\againFiveFemaleMean}{1.76\xspace}
\newcommand{\againSixFemaleMean}{1.91\xspace}
\newcommand{\againSevenFemaleMean}{1.83\xspace}
\newcommand{\likeOneMaleMean}{3.77\xspace}
\newcommand{\likeTwoMaleMean}{3.75\xspace}
\newcommand{\likeThreeMaleMean}{3.7\xspace}
\newcommand{\likeFourMaleMean}{3.74\xspace}
\newcommand{\likeFiveMaleMean}{3.77\xspace}
\newcommand{\likeSixMaleMean}{3.6\xspace}
\newcommand{\likeSevenMaleMean}{3.64\xspace}
\newcommand{\likeOneFemaleMean}{3.78\xspace}
\newcommand{\likeTwoFemaleMean}{3.8\xspace}
\newcommand{\likeThreeFemaleMean}{3.75\xspace}
\newcommand{\likeFourFemaleMean}{3.89\xspace}
\newcommand{\likeFiveFemaleMean}{3.62\xspace}
\newcommand{\likeSixFemaleMean}{3.79\xspace}
\newcommand{\likeSevenFemaleMean}{3.55\xspace}
\newcommand{\againOneTwoMalesMean}{1.93\xspace}
\newcommand{\againTwoTwoMalesMean}{1.89\xspace}
\newcommand{\againThreeTwoMalesMean}{1.93\xspace}
\newcommand{\againFourTwoMalesMean}{1.87\xspace}
\newcommand{\againFiveTwoMalesMean}{1.93\xspace}
\newcommand{\againSixTwoMalesMean}{1.85\xspace}
\newcommand{\againSevenTwoMalesMean}{1.89\xspace}
\newcommand{\againOneTwoFemalesMean}{2\xspace}
\newcommand{\againTwoTwoFemalesMean}{1.97\xspace}
\newcommand{\againThreeTwoFemalesMean}{1.95\xspace}
\newcommand{\againFourTwoFemalesMean}{1.95\xspace}
\newcommand{\againFiveTwoFemalesMean}{1.74\xspace}
\newcommand{\againSixTwoFemalesMean}{1.92\xspace}
\newcommand{\againSevenTwoFemalesMean}{1.81\xspace}
\newcommand{\againOneFemaleMaleMean}{1.88\xspace}
\newcommand{\againTwoFemaleMaleMean}{1.83\xspace}
\newcommand{\againThreeFemaleMaleMean}{1.96\xspace}
\newcommand{\againFourFemaleMaleMean}{2\xspace}
\newcommand{\againFiveFemaleMaleMean}{1.83\xspace}
\newcommand{\againSixFemaleMaleMean}{1.83\xspace}
\newcommand{\againSevenFemaleMaleMean}{1.79\xspace}
\newcommand{\likeOneTwoMalesMean}{3.8\xspace}
\newcommand{\likeTwoTwoMalesMean}{3.76\xspace}
\newcommand{\likeThreeTwoMalesMean}{3.67\xspace}
\newcommand{\likeFourTwoMalesMean}{3.65\xspace}
\newcommand{\likeFiveTwoMalesMean}{3.78\xspace}
\newcommand{\likeSixTwoMalesMean}{3.54\xspace}
\newcommand{\likeSevenTwoMalesMean}{3.63\xspace}
\newcommand{\likeOneTwoFemalesMean}{3.76\xspace}
\newcommand{\likeTwoTwoFemalesMean}{3.79\xspace}
\newcommand{\likeThreeTwoFemalesMean}{3.71\xspace}
\newcommand{\likeFourTwoFemalesMean}{3.92\xspace}
\newcommand{\likeFiveTwoFemalesMean}{3.63\xspace}
\newcommand{\likeSixTwoFemalesMean}{3.89\xspace}
\newcommand{\likeSevenTwoFemalesMean}{3.53\xspace}
\newcommand{\likeOneFemaleMaleMean}{3.75\xspace}
\newcommand{\likeTwoFemaleMaleMean}{3.75\xspace}
\newcommand{\likeThreeFemaleMaleMean}{3.79\xspace}
\newcommand{\likeFourFemaleMaleMean}{3.96\xspace}
\newcommand{\likeFiveFemaleMaleMean}{3.62\xspace}
\newcommand{\likeSixFemaleMaleMean}{3.62\xspace}
\newcommand{\likeSevenFemaleMaleMean}{3.54\xspace}
\newcommand{\heatmapTutoringcomponentDirectInstruction}{ & 10 & 0 & 8 & 1 & 1 & 1 & 10 & 2 \\}
\newcommand{\heatmapTutoringcomponentExplanations}{ & 21 & 7 & 14 & 15 & 11 & 12 & 2 & 2 \\}
\newcommand{\heatmapTutoringcomponentHint}{ & 10 & 2 & 3 & 7 & 7 & 1 & 2 & 0 \\}
\newcommand{\heatmapTutoringcomponentGuidingquestions}{ & 9 & 1 & 1 & 0 & 5 & 2 & 0 & 0 \\}
\newcommand{\heatmapTutoringcomponentOther}{ & 6 & 0 & 4 & 1 & 3 & 0 & 1 & 6 \\}
\newcommand{\heatmapTutoringcomponentSticker}{ & 0 & 38 & 2 & 11 & 8 & 5 & 0 & 0 \\}
\newcommand{\amountFeedbackAllActivities}{215\xspace}
\newcommand{\amountFeedbackActivityOne}{9\xspace}
\newcommand{\amountFeedbackActivityTwo}{14\xspace}
\newcommand{\amountFeedbackActivityThree}{10\xspace}
\newcommand{\amountFeedbackActivityFour}{6\xspace}
\newcommand{\amountFeedbackActivityFive}{35\xspace}
\newcommand{\amountFeedbackActivitySix}{106\xspace}
\newcommand{\amountFeedbackActivitySeven}{35\xspace}
\newcommand{\correlationAgainTaskOne}{0.34\xspace}
\newcommand{\correlationAgainTaskOnePValue}{0.007\xspace}
\newcommand{\correlationAgainTaskTwo}{0.09\xspace}
\newcommand{\correlationAgainTaskTwoPValue}{0.499\xspace}
\newcommand{\correlationAgainTaskThree}{-0.12\xspace}
\newcommand{\correlationAgainTaskThreePValue}{0.346\xspace}
\newcommand{\correlationAgainTaskFour}{0.01\xspace}
\newcommand{\correlationAgainTaskFourPValue}{0.928\xspace}
\newcommand{\correlationAgainTaskFive}{-0.02\xspace}
\newcommand{\correlationAgainTaskFivePValue}{0.854\xspace}
\newcommand{\correlationAgainTaskSix}{-0.04\xspace}
\newcommand{\correlationAgainTaskSixPValue}{0.77\xspace}
\newcommand{\correlationAgainTaskSeven}{-0.02\xspace}
\newcommand{\correlationAgainTaskSevenPValue}{0.895\xspace}
\newcommand{\correlationLikeTaskOne}{0.28\xspace}
\newcommand{\correlationLikeTaskOnePValue}{0.03\xspace}
\newcommand{\correlationLikeTaskTwo}{0.06\xspace}
\newcommand{\correlationLikeTaskTwoPValue}{0.652\xspace}
\newcommand{\correlationLikeTaskThree}{0.2\xspace}
\newcommand{\correlationLikeTaskThreePValue}{0.114\xspace}
\newcommand{\correlationLikeTaskFour}{-0.1\xspace}
\newcommand{\correlationLikeTaskFourPValue}{0.417\xspace}
\newcommand{\correlationLikeTaskFive}{0.11\xspace}
\newcommand{\correlationLikeTaskFivePValue}{0.4\xspace}
\newcommand{\correlationLikeTaskSix}{-0.04\xspace}
\newcommand{\correlationLikeTaskSixPValue}{0.782\xspace}
\newcommand{\correlationLikeTaskSeven}{0.12\xspace}
\newcommand{\correlationLikeTaskSevenPValue}{0.348\xspace}
\newcommand{\correlationAllTasksAgain}{0.06\xspace}
\newcommand{\correlationAllTasksAgainPValue}{0.218\xspace}
\newcommand{\correlationAllTasksLike}{0.11\xspace}
\newcommand{\correlationAllTasksLikePValue}{0.029\xspace}
\newcommand{\amountInteractionsAllTaskspValue}{0.349\xspace}
\newcommand{\frequencyProblemtypeUnderstandingofthetask}{89\xspace}
\newcommand{\meanAgainProblemtypeUnderstandingofthetask}{1.85\xspace}
\newcommand{\meanLikeProblemtypeUnderstandingofthetask}{3.31\xspace}
\newcommand{\proportionWorseAgainProblemtypeUnderstandingofthetask}{13.5\xspace}
\newcommand{\proportionWorseLikeProblemtypeUnderstandingofthetask}{16.9\xspace}
\newcommand{\frequencyProblemtypeWrongcolourcode}{58\xspace}
\newcommand{\meanAgainProblemtypeWrongcolourcode}{1.81\xspace}
\newcommand{\meanLikeProblemtypeWrongcolourcode}{3.59\xspace}
\newcommand{\proportionWorseAgainProblemtypeWrongcolourcode}{17.2\xspace}
\newcommand{\proportionWorseLikeProblemtypeWrongcolourcode}{6.9\xspace}
\newcommand{\frequencyProblemtypeOzobotspezificproblems}{62\xspace}
\newcommand{\meanAgainProblemtypeOzobotspezificproblems}{1.88\xspace}
\newcommand{\meanLikeProblemtypeOzobotspezificproblems}{3.83\xspace}
\newcommand{\proportionWorseAgainProblemtypeOzobotspezificproblems}{11.3\xspace}
\newcommand{\proportionWorseLikeProblemtypeOzobotspezificproblems}{4.8\xspace}
\newcommand{\frequencyProblemtypeReadingdirection}{61\xspace}
\newcommand{\meanAgainProblemtypeReadingdirection}{1.84\xspace}
\newcommand{\meanLikeProblemtypeReadingdirection}{3.71\xspace}
\newcommand{\proportionWorseAgainProblemtypeReadingdirection}{18\xspace}
\newcommand{\proportionWorseLikeProblemtypeReadingdirection}{9.8\xspace}
\newcommand{\frequencyProblemtypeUturn}{48\xspace}
\newcommand{\meanAgainProblemtypeUturn}{1.97\xspace}
\newcommand{\meanLikeProblemtypeUturn}{3.8\xspace}
\newcommand{\proportionWorseAgainProblemtypeUturn}{2.1\xspace}
\newcommand{\proportionWorseLikeProblemtypeUturn}{4.2\xspace}
\newcommand{\frequencyProblemtypeDirection}{50\xspace}
\newcommand{\meanAgainProblemtypeDirection}{1.82\xspace}
\newcommand{\meanLikeProblemtypeDirection}{3.69\xspace}
\newcommand{\proportionWorseAgainProblemtypeDirection}{10\xspace}
\newcommand{\proportionWorseLikeProblemtypeDirection}{8\xspace}
\newcommand{\frequencyProblemtypeApp}{42\xspace}
\newcommand{\meanAgainProblemtypeApp}{1.86\xspace}
\newcommand{\meanLikeProblemtypeApp}{3.32\xspace}
\newcommand{\proportionWorseAgainProblemtypeApp}{11.9\xspace}
\newcommand{\proportionWorseLikeProblemtypeApp}{19\xspace}
\newcommand{\frequencyProblemtypeOther}{7\xspace}
\newcommand{\meanAgainProblemtypeOther}{2\xspace}
\newcommand{\meanLikeProblemtypeOther}{3.86\xspace}
\newcommand{\proportionWorseAgainProblemtypeOther}{0\xspace}
\newcommand{\proportionWorseLikeProblemtypeOther}{0\xspace}
\newcommand{\heatmapOne}{ & 4 & 0 & 5 & 0 & 0 & 0 & 0 & 0 \\}
\newcommand{\heatmapTwo}{ & 6 & 0 & 7 & 0 & 0 & 0 & 0 & 0 \\}
\newcommand{\heatmapThree}{ & 4 & 2 & 1 & 1 & 0 & 0 & 0 & 2 \\}
\newcommand{\heatmapFour}{ & 1 & 4 & 0 & 0 & 0 & 0 & 0 & 0 \\}
\newcommand{\heatmapFive}{ & 6 & 10 & 8 & 4 & 2 & 3 & 0 & 2 \\}
\newcommand{\heatmapSix}{ & 17 & 26 & 8 & 21 & 25 & 10 & 0 & 0 \\}
\newcommand{\heatmapSeven}{ & 9 & 0 & 2 & 0 & 0 & 3 & 15 & 6 \\}
\newcommand{\correlationAgainProblemUnderstanding}{0.03\xspace}
\newcommand{\correlationAgainProblemUnderstandingPValue}{0.517\xspace}
\newcommand{\correlationAgainProblemWrongCode}{0.08\xspace}
\newcommand{\correlationAgainProblemWrongCodePValue}{0.129\xspace}
\newcommand{\correlationAgainProblemSpecific}{-0.01\xspace}
\newcommand{\correlationAgainProblemSpecificPValue}{0.848\xspace}
\newcommand{\correlationAgainProblemReading}{0.04\xspace}
\newcommand{\correlationAgainProblemReadingPValue}{0.391\xspace}
\newcommand{\correlationAgainProblemUturn}{-0.1\xspace}
\newcommand{\correlationAgainProblemUturnPValue}{0.05\xspace}
\newcommand{\correlationAgainProblemDirection}{0.06\xspace}
\newcommand{\correlationAgainProblemDirectionPValue}{0.235\xspace}
\newcommand{\correlationAgainProblemApp}{0.02\xspace}
\newcommand{\correlationAgainProblemAppPValue}{0.724\xspace}
\newcommand{\correlationAgainProblemOther}{-0.05\xspace}
\newcommand{\correlationAgainProblemOtherPValue}{0.345\xspace}
\newcommand{\correlationLikeProblemUnderstanding}{0.18\xspace}
\newcommand{\correlationLikeProblemUnderstandingPValue}{.001\xspace}
\newcommand{\correlationLikeProblemWrongCode}{0\xspace}
\newcommand{\correlationLikeProblemWrongCodePValue}{0.968\xspace}
\newcommand{\correlationLikeProblemSpecific}{-0.12\xspace}
\newcommand{\correlationLikeProblemSpecificPValue}{0.014\xspace}
\newcommand{\correlationLikeProblemReading}{-0.06\xspace}
\newcommand{\correlationLikeProblemReadingPValue}{0.209\xspace}
\newcommand{\correlationLikeProblemUturn}{-0.1\xspace}
\newcommand{\correlationLikeProblemUturnPValue}{0.052\xspace}
\newcommand{\correlationLikeProblemDirection}{-0.05\xspace}
\newcommand{\correlationLikeProblemDirectionPValue}{0.349\xspace}
\newcommand{\correlationLikeProblemApp}{0.11\xspace}
\newcommand{\correlationLikeProblemAppPValue}{0.022\xspace}
\newcommand{\correlationLikeProblemOther}{-0.04\xspace}
\newcommand{\correlationLikeProblemOtherPValue}{0.376\xspace}
\newcommand{\frequencyTutoringcomponentDirectInstruction}{74\xspace}
\newcommand{\meanAgainTutoringcomponentDirectInstruction}{1.84\xspace}
\newcommand{\meanLikeTutoringcomponentDirectInstruction}{3.37\xspace}
\newcommand{\frequencyTutoringcomponentExplanations}{167\xspace}
\newcommand{\meanAgainTutoringcomponentExplanations}{1.86\xspace}
\newcommand{\meanLikeTutoringcomponentExplanations}{3.58\xspace}
\newcommand{\frequencyTutoringcomponentHint}{63\xspace}
\newcommand{\meanAgainTutoringcomponentHint}{1.94\xspace}
\newcommand{\meanLikeTutoringcomponentHint}{3.9\xspace}
\newcommand{\frequencyTutoringcomponentGuidingquestions}{31\xspace}
\newcommand{\meanAgainTutoringcomponentGuidingquestions}{1.9\xspace}
\newcommand{\meanLikeTutoringcomponentGuidingquestions}{3.37\xspace}
\newcommand{\frequencyTutoringcomponentOther}{50\xspace}
\newcommand{\meanAgainTutoringcomponentOther}{1.94\xspace}
\newcommand{\meanLikeTutoringcomponentOther}{3.65\xspace}
\newcommand{\frequencyTutoringcomponentSticker}{97\xspace}
\newcommand{\meanAgainTutoringcomponentSticker}{1.83\xspace}
\newcommand{\meanLikeTutoringcomponentSticker}{3.63\xspace}
\newcommand{\correlationAgainTutoringDirectInstruction}{0.05\xspace}
\newcommand{\correlationAgainTutoringDirectInstructionPValue}{0.311\xspace}
\newcommand{\correlationAgainTutoringExplanations}{0.05\xspace}
\newcommand{\correlationAgainTutoringExplanationsPValue}{0.361\xspace}
\newcommand{\correlationAgainTutoringHint}{-0.07\xspace}
\newcommand{\correlationAgainTutoringHintPValue}{0.133\xspace}
\newcommand{\correlationAgainTutoringGuiding}{-0.02\xspace}
\newcommand{\correlationAgainTutoringGuidingPValue}{0.646\xspace}
\newcommand{\correlationAgainTutoringOther}{-0.07\xspace}
\newcommand{\correlationAgainTutoringOtherPValue}{0.165\xspace}
\newcommand{\correlationAgainTutoringSticker}{0.07\xspace}
\newcommand{\correlationAgainTutoringStickerPValue}{0.19\xspace}
\newcommand{\correlationLikeTutoringDirectInstruction}{0.13\xspace}
\newcommand{\correlationLikeTutoringDirectInstructionPValue}{0.01\xspace}
\newcommand{\correlationLikeTutoringExplanations}{0.01\xspace}
\newcommand{\correlationLikeTutoringExplanationsPValue}{0.881\xspace}
\newcommand{\correlationLikeTutoringHint}{-0.17\xspace}
\newcommand{\correlationLikeTutoringHintPValue}{.001\xspace}
\newcommand{\correlationLikeTutoringGuiding}{0.08\xspace}
\newcommand{\correlationLikeTutoringGuidingPValue}{0.114\xspace}
\newcommand{\correlationLikeTutoringOther}{-0.03\xspace}
\newcommand{\correlationLikeTutoringOtherPValue}{0.574\xspace}
\newcommand{\correlationLikeTutoringSticker}{-0.03\xspace}
\newcommand{\correlationLikeTutoringStickerPValue}{0.546\xspace}

%% file: content/introduction.tex
\section{Introduction}
\label{sec:introduction}

Computer science related topics are increasingly taught in primary schools around the world~\cite{heintz2016}, and programming is commonly used as a vehicle to promote different aspects of computational thinking~\cite{mannila2014computational}.
This, however, challenges primary school teachers~\cite{greifenstein2021challenging}, who have to cover many subjects without specialising in all of them, such that they often lack subject knowledge~\cite{sentance2017computing}.
This problem is further exacerbated by the ongoing debate about the appropriate approach for teaching programming in primary schools: Programming concepts can be taught unplugged, using computers, or with physical approaches using programmable robots and microcontrollers. Even within the specific domain of programming with robotics, new educational robots emerge regularly~\cite{sullivan2016,catlin2018edurobot}.

The lack of appropriate subject knowledge particularly affects the ability to
provide feedback~\cite{greifenstein2021challenging} and to assist when learners
face challenges such as debugging their
programs~\cite{yadav2016expanding,michaeli2019improving}. Especially for giving
corrective and meaningful feedback, content knowledge as well as pedagogical
content knowledge are needed: Elaborated feedback includes, e.g., explanations
of concepts or hints on how to proceed~\cite{narciss2013designing}, for which a
prerequisite is to understand the root cause of the problem experienced by the
learner; in other words, a teacher needs to be able to debug the problem on the
fly.
Consequently, one aim of this paper is to shed light on what are common
problems encountered in primary school programming activities, in order to
adequately prepare teachers to provide feedback on them.

While corrective feedback on problems is crucial in the learning process, it however might interfere with the experienced fun: Fun is strongly related to intrinsic motivation~\cite{malone2021making} and intrinsic motivation can be decreased by corrective feedback~\cite{wisniewski2020power}. 
This is problematic, as (1) intrinsic motivation is needed to develop
individual interest and to pursue a learning goal independent of
extrinsic rewards~\cite{ryan2000intrinsic,wisniewski2020power}, and
(2) motivation and interest are considered some of the most important
aims of programming in primary schools~\cite{greifenstein2021challenging} in general.
Moreover, fun activities can lead to situational interest that represents the earlier phases of interest development~\cite{renninger2009interest}.
As a consequence, motivating approaches such as game-based and active
learning~\cite{straubinger2022gamekins,anewalt2008making} are commonly
used in the primary school
classroom~\cite{hainey2016systematic}. 
Since giving students corrective feedback might affect the fun the students experience, the second aim of this paper is to understand whether the different problems and different tutoring types of feedback have an effect on fun.

As introducing children to programming early can reduce or even avoid
prejudices and fears~\cite{sullivan2016,vrieler2020}, it is a common strategy
to encourage girls in particular to become enthusiastic about
programming~\cite{sullivan2016,aivaloglou2019a}. This is expected to increase
their self-efficacy and STEM-related confidence~\cite{bosu2019,albusays2021}
while overcoming their anxiety of the perceived male dominated computer science
and associated negative stereotypes~\cite{lishinski2016,aivaloglou2019a}. In order to support these efforts, it is important to understand how the problems girls experience affect them. Therefore, a third aim of this paper is to investigate whether any gender-specific differences can be observed with respect to problems and fun experienced.

In this paper, we evaluate the combination of corrective feedback on common problems and encouraging students via Ozobot Evo programming activities:

\vspace{0.5em}
\noindent\textbf{RQ 1:} \rqOne

\vspace{0.5em}
\noindent\textbf{RQ 2:} \rqTwo

\vspace{0.5em}
\noindent\textbf{RQ 3:} \rqThree
\vspace{0.5em}

To answer these research questions, we conducted a study consisting of seven workshops with \numParticipantsThirdGrade third and \numParticipantsFourthGrade fourth grade students. They were tasked to perform seven programming activities with the Ozobot Evo robot. The students noted their experienced fun and we noted the students' problems and our feedback. To support teachers with giving corrective feedback in an encouraging learning climate
, we summarise the students' problems and our ideas for appropriate and efficient feedback.

Our results indicate that feedback should, whenever possible, include
the participation of the students, for example by giving hints without
revealing the solution. Moreover, feedback on urgent problems that
need to be solved quickly to complete the task, or problems that might
be attributed to personal abilities (and thus might decrease
self-efficacy) tend to be more discouraging than feedback on other
problems.  This is why we suggest that especially task constraints or
difficult concepts should be clearly addressed in front of the class
or adequately prepared in additional material.

%% file: content/relatedwork.tex
\section{Related Work}
\label{sec:relatedwork}

The context of our study is primary school education, where the introduction of programming concepts is frequently conducted using programmable robots such as the Ozobot Evo.

\subsection{Educational Use of Robots}
The educational use of microcontrollers and robots in general has been shown to enhance learning and engage students in STEM fields~\cite{benitti2012exploring,kubilinskiene2017applying,przybylla2018impact,conde2021fostering,anwar2019systematic}. Regarding primary school teachers, the results are rather mixed~\cite{lathifah2019contribution}: They are significantly less enthusiastic about using educational robots than secondary school teachers~\cite{reich2016robots}. This might be attributed to their low confidence and lacking knowledge observed at the beginning of workshops on robots: both have been observed to increase throughout the course of such workshop~\cite{kay2012using,kim2015robotics}. 

\subsection{Teacher Training on Robotics}

Generally, teacher training is the most common strategy of experienced teachers to master teacher-related challenges~\cite{greifenstein2021challenging}. For researchers and in particular teacher trainers, the question on how to design a workshop for teachers emerges. Experienced teachers prefer teacher training that promotes both content knowledge and pedagogical content knowledge (PCK)~\cite{greifenstein2021challenging}. Regarding the PCK, supporting individual students is considered a main challenge that is further exacerbated by the lack of time, diagnostic skills of teachers, open-ended nature of programming and ratio of students to teachers~\cite{michaeli2019current,sentance2017computing,yadav2016expanding}. Teachers should therefore be supported with dealing with common problems by providing appropriate teacher training or ideas for auxiliary material for the students.

\subsection{Robots and Feedback}

There are many studies on the effects of different robots in the
primary school classroom such as LEGO WeDo~\cite{mayerove2017teach},
the mBot~\cite{pisarov2019programming}, or
Bee-Bot~\cite{cacco2014bee}. Recent studies focused, e.g., on feedback
regarding the productive collaboration between
students~\cite{zakaria2022} or a gender-neutral design of the
robot~\cite{sullivan2016}. However, only few studies focused on the
problems experienced while introducing robots, e.g., with the Arduino
programming syntax, the malfunction of digital pins~\cite{zhong2020}
or the controllability of the Thymio II
robot~\cite{riedo2013thymio}.
To the best of our knowledge, there is no prior work explicitly
focusing on how to support teachers with giving feedback on the
programming activities, neither for Ozobot robots, nor for others.
There are, however, several experience reports~\cite{korber2021experience,benvenuti2019using,tengler2020klein,picka2020robotic,fojtik2017ozobot,mayerova2019creating,chou2018little,vzavcek2019development,van2018best}: While these reports
mention some flaws, they nevertheless clearly recommend the use of
Ozobots because of its positive
effects. This
is why we aim at bridging these two aspects by measuring the
experienced fun, but also investigating the effects of common problems
and their associated feedback.

\subsection{Effective Feedback and Fun}
For feedback to be effective, it has to fulfil certain criteria such as being timely and actionable~\cite{wiggins2012seven}. The latter implies that giving praise or grades does not suffice to support students in improving~\cite{wisniewski2020power}. Instead, elaborated feedback should be given, e.g., in terms of hints, explanations or guiding questions~\cite{narciss2013designing}.
Generally, corrective feedback might at the same time hamper learning as it can impair motivational aspects~\cite{wisniewski2020power}. This is crucial, as less motivated learners process corrective feedback worse~\cite{depasque2015effects}.
To improve the processing of corrective feedback, one common assumption is that the students should enjoy the activities. Motivation and fun are correlated as ``intrinsic motivation is defined as the doing of an activity for its inherent satisfactions [...]''~\cite{ryan2000intrinsic} and ``when intrinsically motivated a person is moved to act for the fun or challenge entailed [...]''~\cite{ryan2000intrinsic}. Fun activities have also been shown to be an effective intrinsic motivator in programming education~\cite{long2007just,riedo2013thymio}. Positive emotions during learning activities raise situational interest, which in turn can lead to individual interest in the long term~\cite{renninger2009interest} as ``individual interest is characterized by positive emotions, such as enjoyment, personal meaningfulness, knowledge, and continued engagement over time''~\cite{renninger2019cambridge}.
To strike a balance between corrective feedback and fun, thus to enable learning but at the same time reaching the goal of getting children interested in programming~\cite{greifenstein2021challenging}, feedback should also be given in an encouraging way. While there are profound findings on the encouraging effects of positive feedback~\cite{hattie2008visible,depasque2015effects}, in this paper, we look at ways how to give corrective feedback in an encouraging way.

\subsection{Gender in Robotics}
When it comes to increasing the ratio of women in STEM disciplines, it is relevant to understand how children learn to program with robots and thus address potential gender-specific behaviours in the most effective way. 
In the field of robot programming, boys have been reported to perform better than girls in more advanced tasks such as repeat-loops with KIWIE robotics kit, while no difference was found in basic tasks~\cite{sullivan2013, sullivan2016}. This may be a result of boys having more exposure to conventional ``male'' toys, and thus are more interested in robotics and take more risks in programming than girls~\cite{sullivan2016,vrieler2020}.
In addition, girls and boys have different learning approaches, methods and expectations in programming; for example, girls prefer discussing and debating in groups more than boys~\cite{rubio2015a, papavlasopoulou2020,vrieler2020}.
In this context, pair programming can help to increase enthusiasm and learning outcomes in programming in primary school~\cite{iskrenovic-momcilovic2019}.
In learning groups, it has been shown that students' grades are slightly better in same-sex groups than in mixed groups~\cite{zeid2011} and all-female groups are more engaged with tasks and follow instructions better~\cite{mcdowell2006, zhan2015, sullivan2016}. Overall, quick positive feedback in tasks may result in a higher engagement of girls in programming activities~\cite{vrieler2020}.

%% file: content/workshop.tex
\section{The Ozobot Programming Workshop}
\label{sec:workshop}

In order to enable our investigations, we designed a workshop that introduces primary school children to programming using the popular Ozobot Evo robots.

\begin{figure}[t]
	\centering
	\subfloat[\label{fig:ozobotpath}
	Ozobot Evo's path of activity~3.]{\includegraphics[width=0.48\columnwidth]{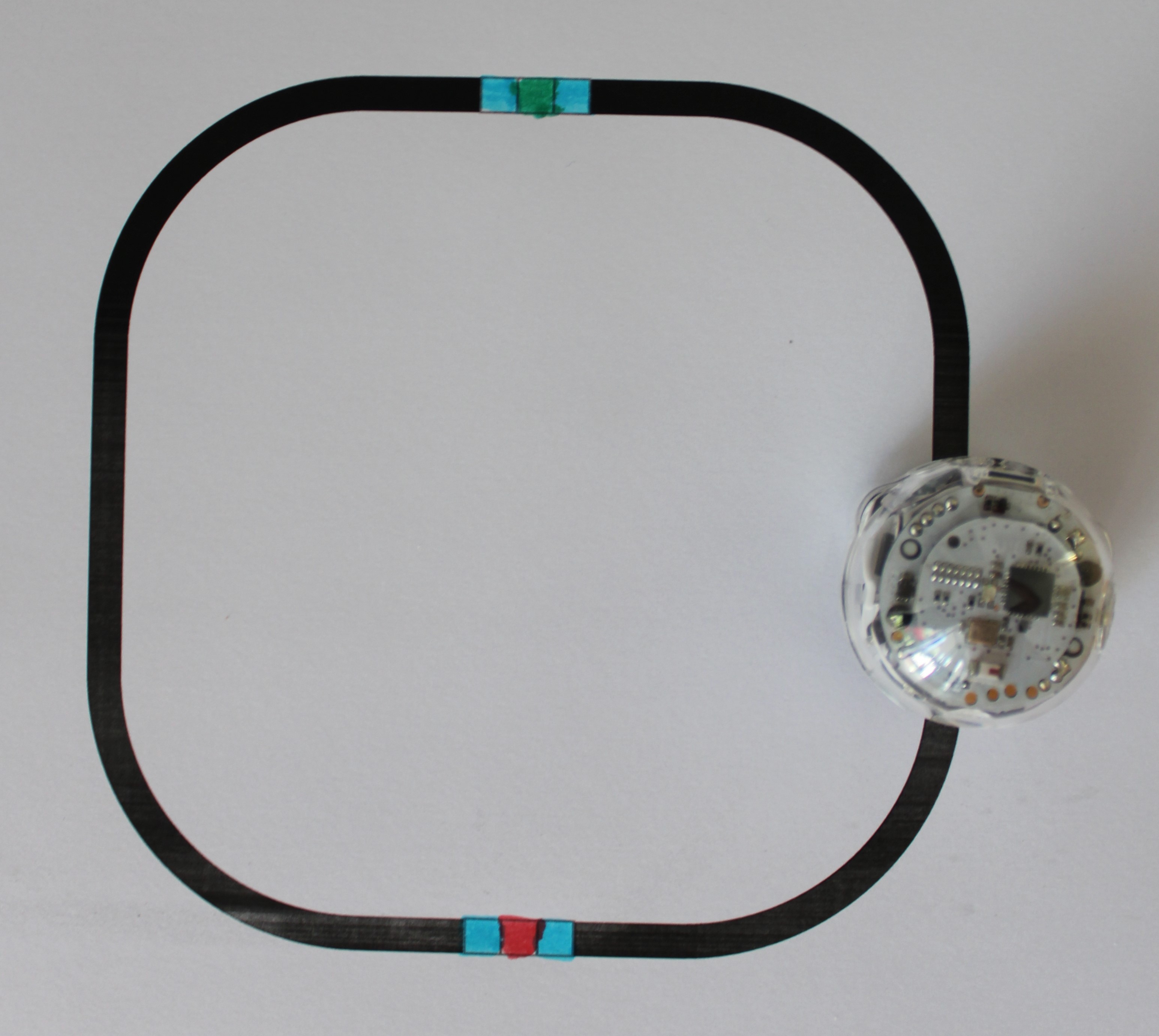}}\hspace{0.1em}
	\subfloat[\label{fig:ozoblockly}OzoBlockly code for activity~7.]{\includegraphics[width=0.48\columnwidth]{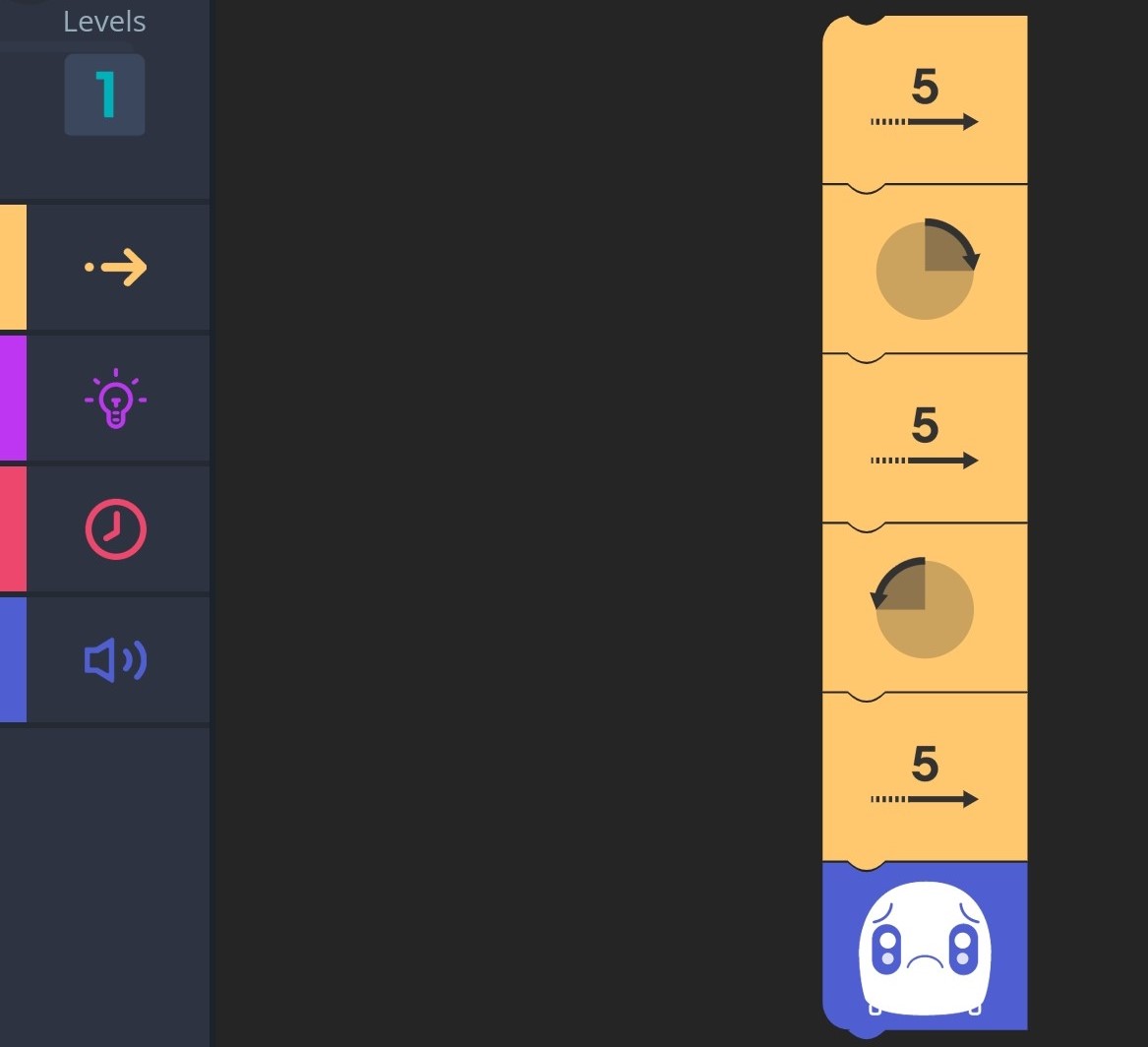}}
	\vspace{-1em}
	\caption{The two programming modes.} 
	\label{fig:modes}
\end{figure}

\input{tables/overview}

\subsection{The Ozobot Evo Robot}

We decided to use the Ozobot Evo robot (see \cref{fig:ozobotpath}) especially because it can be programmed in two ways. In both modes, the Ozobot Evo's actuators and sensors enable sounds, flashing, and moving in various ways such as zigzagging or U-turns.

\subsubsection{The Pen-and-paper Mode}
In the pen-and-paper mode, the Ozobot Evo can be controlled via colours and colour codes (see \cref{fig:ozobotpath}). It follows coloured lines and its lights change according to the colour of the line. The colour codes consist of two or three small coloured segments that are inserted into a line and offer more possibilities: There are predefined colour codes for controlling the speed, direction, finish, special moves, timers and counters\footnote{https://play.ozobot.com/print/guides/ozobot-ozocodes-reference.pdf, last accessed June 3, 2022}.

\subsubsection{Programming with Digital Devices}
The Ozobot Evo can also be programmed via the app ``Ozobot Evo'' or a
web interface\footnote{https://ozobot.com/create/ozoblockly, last
  accessed June 3, 2022}. Both options include the block-based
programming language
``OzoBlockly''. 
The blocks are separated into categories and levels and can
be drag-and-dropped to implement a program (see
\cref{fig:ozoblockly}). While we mainly used level 1 which contains
few and only graphical blocks with numbers, in the levels 2--5 many
textual blocks and further categories such as ``Functions'' and
``List'' are available. The programs of the app are transferred to the
Ozobot Evo via Bluetooth. 

\subsection{Ozobot Evo Activities}
\label{sec:activities}

The workshop is designed for a duration of 90 minutes, or roughly two lessons. During this time, the students perform seven Ozobot Evo activities in the procedure suggested by Körber et al.~\cite{korber2021experience} and summarised in \cref{fig:overview}. These activities are based on teaching material available online and consist of six pen-and-paper mode activities\footnote{https://storage.googleapis.com/ozobot-lesson-library/3-5-basic-training-color-codes/3-5-Basic-Training-Color-Codes-full-version.pdf, last accessed June 3, 2022} and one digital mode activity\footnote{https://storage.googleapis.com/ozobot-lesson-library/ozoblockly-training-k-1/ozoblockly-training-k-1.pdf, last accessed June 3, 2022}. Whenever a student or group of students completes an activity early, they can spend some time drawing their own lines for the robot to follow. 

\subsection{Learning Objectives}

The workshops of 90 minutes each aimed at the students...
\begin{itemize}
\item getting to know sensors and actuators of a robot,
\item using abstraction to match desired behaviour of the robot and available colour codes,
\item observing the robot while executing a path in a specific order and loops in circles,
\item debugging a given sequence in a block-based environment,
\item collaborating with each other to develop a solution,
\item and moreover, enjoying their first programming experience.
\end{itemize}

The children first elaborated on each activity in pairs and to consolidate the learning objectives, we discussed the children's observations in class (see \cref{fig:overview}).

%% file: tables/overview.tex

%
%
%
%
%


\begin{figure*}
\begin{minipage}[c]{\textwidth}


\framebox{\parbox[b][0.25cm][c]{0.30\textwidth}{\textbf{0: Introduction}}
} 
\framebox{\parbox[b][0.25cm][c]{0.36\textwidth}{\textbf{1+2: Line-Following}}
}
\framebox{\parbox[b][0.25cm][c]{0.27\textwidth}{\textbf{3+4: Commands}}
}


\framebox{\parbox[t]{0.30\textwidth}{
\faCommenting \space Collecting prior knowledge on robots \\
\faUser \faUser \space Exploring equipment of Ozobot Evo \\
\faCommenting \space Collecting findings}
} 
\framebox{\parbox[t]{0.36\textwidth}{
\faUser \faUser \space Inserting black (worksheet 1) and coloured lines (worksheet 2) \\
\faCommenting \space Reflecting on the colour sensor}
}
\framebox{\parbox[t]{0.27\textwidth}{
\faUser \faUser \space Inserting given colour codes (worksheet 3 and 4) \\
\faCommenting \space Reflecting on the symmetry}
}
\end{minipage}\\

\begin{minipage}[c]{\textwidth}


\framebox{\parbox[b][0.25cm][c]{0.40\textwidth}{\textbf{5: Randomness}}
}
\framebox{\parbox[b][0.25cm][c]{0.15\textwidth}{\textbf{6: Controlling}}
}
\framebox{\parbox[b][0.25cm][c]{0.38\textwidth}{\textbf{7: OzoBlockly}}
}


\framebox{\parbox[t]{0.40\textwidth}{
\faUser \faUser \space Tracking the arrival colour (worksheet 5/left) \\
\faCommenting \space Collecting the data \\
\faUser \faUser \space Inserting a code from the table (worksheet 5/right)}
}
\framebox{\parbox[t]{0.15\textwidth}{
\faUser \faUser \space Inserting suitable colour codes (worksheet 6)
}
}
\framebox{\parbox[t]{0.38\textwidth}{
\faCommenting \space Executing sequences using printed commands
\\
\faCommenting \space Demonstrating the OzoBlockly app \\
\faUser \faUser \space Debugging an erroneous program (worksheet 7) }
}

\vspace{1em}

\centering \faUser \faUser \space = working in pairs; \faCommenting \space = class discussion
\end{minipage}
\vspace{-0.5em}
\caption{\label{fig:overview} Overview of the Programming Workshop Schedule described in \cref{sec:activities}.}
\end{figure*}

%% file: content/method.tex
\section{Research Method}
\label{sec:method}

\input{figures/notesheet}

Using the Ozobot Evo workshop, we aim to  empirically answer our research questions specified at the end of \cref{sec:introduction}.

\subsection{Participants}
\looseness=-1
We conducted \numWorkshops workshops with a total of \numTotalParticipants  children aged eight to ten years; \numParticipantsThirdGrade students were in the third grade, and \numParticipantsFourthGrade in the fourth grade. In total, there were \numGirls girls, and \numBoys boys.
Whenever possible, we split them into groups of two (mostly the seatmates). This resulted in \numTwoFemalesGroups all-female, \numTwoMalesGroups all-male, \numFemaleMaleGroups mixed groups (consisting of one girl and one boy), and \numOneStudentGroups students worked individually.

\subsection{Data Collection}

In every workshop, three to four supervisors were available to answer students' questions. In total, there were two male and six female supervisors, including three researchers and five undergraduates in the field of computer science and teaching studies. None of them knew the children before which may result in less individualised but therefore also more neutral feedback as if the feedback was given by the teacher.

For each activity (\cref{fig:overview}), the children first received an explanation and then performed it. When they encountered a problem they raised a hand, and one of us then gave feedback to this group until they overcame their difficulties. Immediately after each conversation we took a note of it on the sheet shown in \cref{fig:notesheet}.
After each activity (\cref{fig:overview}), each child filled in the ``Fun Toolkit'', which measures fun with children efficiently, as it requires only little writing and reading~\cite{read2008validating}. We implemented the Fun Toolkit using the ``Smileyometer'' (\cref{fig:smileyometer}) and the ``Again Again Table'' (\cref{fig:againagain}).

\subsection{Data Analysis}

In order to answer the three main research questions, we consider the two sources of data: The supervisor notes on problems and feedback, and the student perception measured with the Fun Toolkit.

\subsubsection{Students' Problems and Supervisors' Feedback}

The notes on the students' problems and supervisors' feedback provided us with qualitative data on which we applied thematic analysis~\cite{bergman2010hermeneutic}. We collected themes, then counted them and in a final step again related them to the original data and our research questions. To ensure inter-rater reliability ($K$ = \irr), two authors independently annotated all qualitative data.
To categorise the problems that the students faced (RQ1), we used the third column of the notes (\cref{fig:notesheet}) where the supervisor noted the students' statements and the actual issue. 
To answer RQ2 we additionally categorised the notes on the supervisors' feedback (fourth column of \cref{fig:notesheet}) extending the tutoring strategies by Narciss~\cite{narciss2013designing}.

\subsubsection{Correlation of Supervisors' Feedback and Students' Fun}
The ``Again Again Table'' and the ``Smileyometer'' (\cref{fig:funToolkit}) both contain scales that have to be filled for each activity summarised in \cref{fig:overview}. This provides us with two values between 0 and 4, respectively 0 and 2, per child and activity. We compared the means to find correlations between being given specific feedback and the experienced fun. Moreover, the values of the used point-biserial correlation coefficient $r_{pb}$ can range from -1 to 1---indicating the most negative (-1) to no (0) to the most positive correlation (1).

\subsubsection{Gender Differences}
Besides the overall classification of problems, we further aim to determine if
the group constellations have an influence on the occurrence of these problems
(RQ3). After the thematic analysis, we therefore used a Kruskal-Wallis H test
to measure significant differences between the three group constellations
(female only, male only, mixed). We calculated $\eta^{2}$ to measure the effect
size. If $\eta^{2} > 0.14$ there are large, $\eta^{2} > 0.06$ there are medium
and $\eta^{2} > 0.01$ there are small effects.
We also measured statistical differences between gender and between the group constellations using a Wilcoxon Rank Sum test with $\alpha = 0.05$ and calculated $r$ to measure the effect size. If $r > 0.5$ there are large effects, $r > 0.3$ there are medium and $r > 0.1$ there are small effects.

\subsection{Limitations}

The limitations mostly result from the workshop design, which is one
specific workshop which lasts only 90 minutes. However, even short
lessons can increase interest in
programming~\cite{relkin2020}. Teaching programming with Ozobot robots
for a longer time might reveal further recurring problems as well as
problems to be solved only once.  Furthermore, our workshop focused on
the pen-and-paper-mode (activities~1 to 6) but in subsequent lessons,
more block-based programming activities could follow activity~7. This
might shift the focus of the problems from robot-specific content to
algorithmic control structures. Besides the problems, it might also be
interesting to see how the fun develops over time and if
(not) using the pen-and-paper mode in introductory lessons affects the
experienced fun of further lessons.

\input{figures/funToolkit}

%% file: figures/notesheet.tex
\begin{figure}[tb]
	\includegraphics[width=\columnwidth]{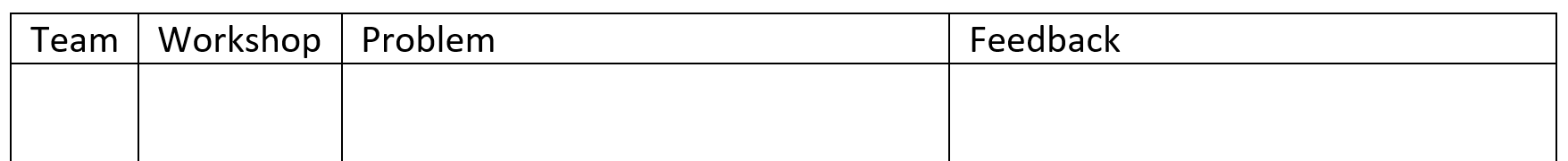}
	\caption{\label{fig:notesheet}Columns of the sheet used to take notes.}
	\vspace{-1em}
\end{figure}

%% file: figures/funToolkit.tex
\begin{figure}[t]
	\centering
	\subfloat[\label{fig:smileyometer}The Smileyometer.]{\includegraphics[width=0.64\columnwidth]{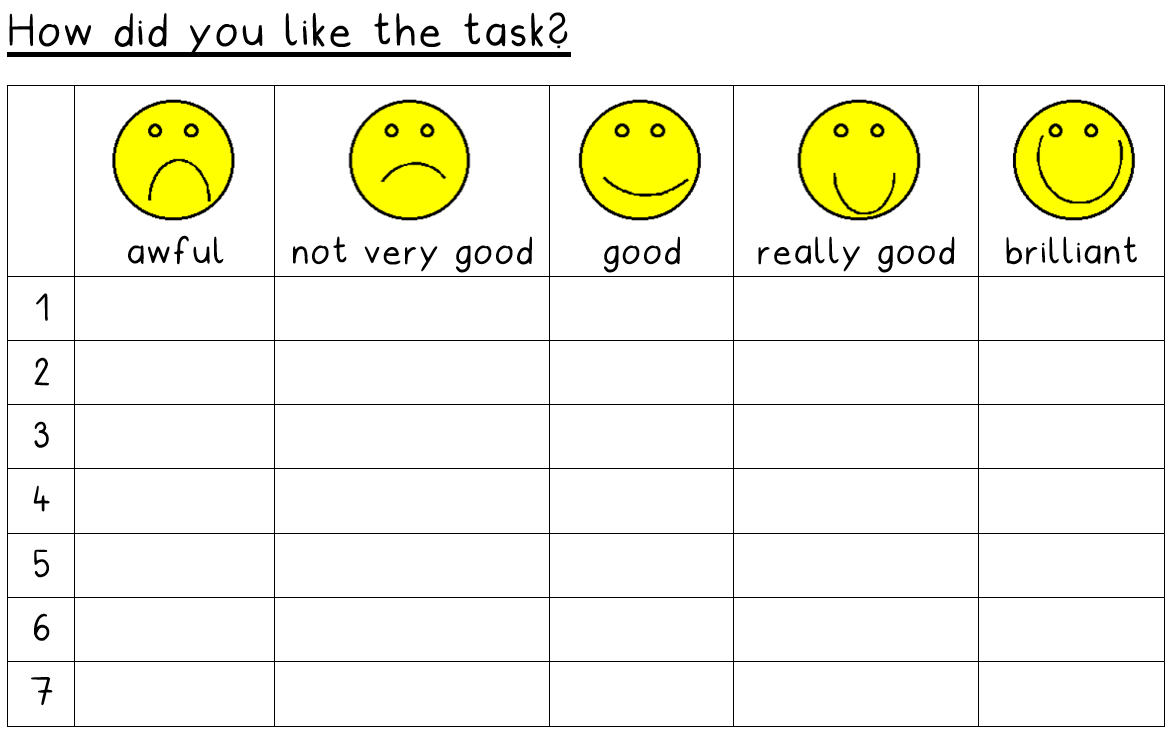}}
	\hspace{0.1em}
	\subfloat[\label{fig:againagain}The Again Again table.]{\includegraphics[width=0.34\columnwidth]{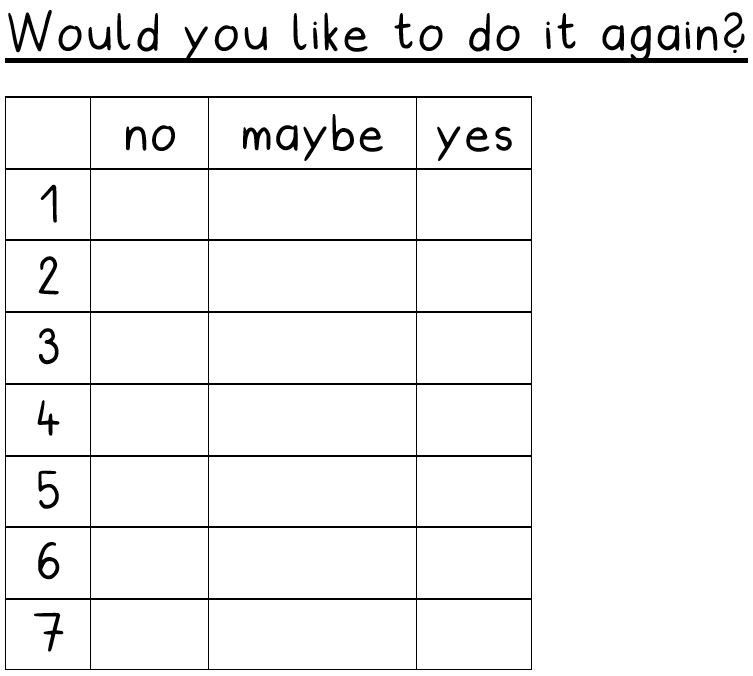}}
	\vspace{-1em}
	\caption{The Fun Toolkit used to measure experienced fun.} 
	\label{fig:funToolkit}
\end{figure}

%% file: content/results.tex
\section{Results}
\label{sec:results}

Using the data collected throughout the seven workshops, we empirically answer our
three main research questions.

\subsection{RQ 1: Problems and Support}
\input{tables/heatmapProblemsActivities}

\input{tables/heatmapProblemsFeedback}

To answer RQ1 we consider the supervisors' notes (\cref{fig:notesheet}).
\Cref{tab:heatmapProblemsActivities} and \cref{tab:heatmapProblemsFeedback} summarise the problems encountered by the
primary school children, and relate them to the activities respectively the corresponding tutoring feedback type given.
The categorisation of the feedback resulted in six different categories that are inspired by Narciss' tutoring feedback strategies~\cite{narciss2013designing} and listed by their frequency:
\begin{itemize}
	\item \textbf{T1: Explanations} describe general clarifications on the task or the robot. 
	\item \textbf{T2: Stickers} refer to problems where a student error required white stickers to hide the mistake and let the students re-do the task.
	\item \textbf{T3: Direct Instruction} refers to information on specific steps to achieve a certain task, for example when using the app.
	\item \textbf{T4: Hints} describe suggestions that indicate how to complete the task.
	\item \textbf{T5: Other} refers to feedback not matching any of the above categories.
	\item \textbf{T6: Guiding questions} can be used to help students arrive at a solution on their own.
\end{itemize}

The categorisation of problems resulted in eight distinct categories (P1--P8),
which are discussed in detail in the following in the order of their frequency starting with the most frequent,
together with the forms of feedback used. The problems are discussed in the
context of the associated activities that are
summarised in \cref{fig:overview}. 

\subsubsection{P1: Understanding of the Activity}
The most frequent problems encountered by primary school children deal with understanding the activity. For all activities, some students of the workshops had difficulties understanding the task constraints (P1, \cref{tab:heatmapProblemsActivities}). It often was sufficient to give an explanation, hints or guiding questions to the students as feedback (T1, T4 and T6, \cref{tab:heatmapProblemsFeedback}). In case of major comprehension problems, direct instruction was needed (T3, \cref{tab:heatmapProblemsFeedback}).

\subsubsection{P2: Wrong Colour Code}
Wrong colour codes occur not only because of misinterpreted directions or U-turns (P5 and P6) but also by accidentally copying codes incorrectly---especially for activity~6 where five different codes had to be inserted (P2, \cref{tab:heatmapProblemsActivities}). To enable the students to proceed, a new worksheet---especially if there are many mistakes---or a sticker can be given (T2, \cref{tab:heatmapProblemsFeedback}).

\subsubsection{P3: Ozobot Evo Specific Aspects}
Another common problem deals with the characteristics of the Ozobot Evo such as its sensors and actuators. As can be seen in \cref{tab:heatmapProblemsFeedback} (T1), giving explanations on these concepts is the most frequently chosen kind of feedback. Direct instruction on how to deal with the Ozobot Evo's characteristics was also necessary sometimes (T3, \cref{tab:heatmapProblemsFeedback}).
In activities~5 and 6, the children had to choose colour codes on their own for the first time (P3, \cref{tab:heatmapProblemsActivities}). This led to some groups being confused by one colour code described as ``jumping''.
Another frequent issue when dealing with colour codes relates to drawing a custom line that was also experienced in another workshop~\cite{french2018using}: It is important to draw a long enough single-coloured line before and after every code to ensure that the Ozobot Evo's sensors recognise the respective colour code.

\subsubsection{P4: Reading Directions}
Students often recognised the issue of the reading directions on their own stating ``In what order do we have to draw the colours?'' or ``We should have started with green''. The support, however, differs: In the first case, the teacher should explain the concept and in the latter, the teacher just has to hand out a sticker to cover the wrong code (T1 and T2, \cref{tab:heatmapProblemsFeedback}). While the codes could mostly be drawn in the same order as specified, in activity~6, the students had to put themselves into the perspective of the robot: If the robot arrives at the code from the right side, the code has to be drawn the other way round which made activity~6 complex regarding spatial thinking (P4, \cref{tab:heatmapProblemsActivities}).

\subsubsection{P5: U-turns}
Also for activity~6, U-turns were introduced---which seem to be difficult to understand, also because there are two colour codes for U-turns (P5, \cref{tab:heatmapProblemsActivities}). In most interactions regarding U-turns, explanations were given, accompanied by white stickers to correct errors where necessary (T1 and T2, \cref{tab:heatmapProblemsFeedback}). This enables students to cover the wrong code and draw a new colour code. 

\subsubsection{P6: Direction}
Turning or moving into a specific direction was an issue in activities 5 to 7 (P6, \cref{tab:heatmapProblemsActivities}). However, students saying, e.g., ``It's not going to the left'' could require different explanations depending on the misconceptions of the students (T1, \cref{tab:heatmapProblemsFeedback}).
While some groups simply confused left and right, others did not mind that the Ozobot Evo turns randomly at crossings if there is no corresponding colour code. These issues were solved with explanations of the directions or the randomness. When codes were already drawn, stickers also have a supporting effect (T2, \cref{tab:heatmapProblemsFeedback}).

\subsubsection{P7: App}
During activity~7, we demonstrated the relevant features of the OzoBlockly app (P7, \cref{tab:heatmapProblemsActivities}). However, some groups needed further direct instruction on deleting blocks and running the program, or information on programming sequences (T3, \cref{tab:heatmapProblemsFeedback}). 

\subsubsection{P8: Other Aspects}
There are further technical and organisational aspects such as the Ozobot Evo not driving anymore which is not due to a conceptual understanding of the students, but for other reasons such as the Ozobot Evo not being turned on or an empty battery (which happens frequently in the digital mode~\cite{mayerova2019creating}). This comparatively trivial problem can best be clarified quickly by explanations or by replacing the Ozobot Evo (T1 and T5, \cref{tab:heatmapProblemsFeedback}).

\vspace{1em}
\summary{RQ1}{The eight common problems can be generalised to different degrees: For other learning robots, it might also be important to ensure a general understanding of the activity and the app and have explanations for directional movements. When programming the Ozobot Evo
, many common problems can be solved by providing white stickers and spare robots.}

\subsection{RQ 2: Feedback and Fun}

\input{figures/boxplotAgain}
\input{figures/boxplotLike}

To answer RQ2 we consider the data of the Again Again table and the Smileyometer. \cref{fig:boxplotAgain} and \cref{fig:boxplotSmileyometer} show the distribution regarding the activities that are summarised in \cref{fig:overview}
.


\subsubsection{Measured Fun}
Generally, the values of fun are very high with a mean of \allActivitiesAgainMean for the Again Again table and \allActivitiesLikeMean for the Smileyometer. These values might be considered when looking at \cref{tab:problemsFun}. Generally, most students wanted to perform the respective activity again and considered it ``brilliant'' or ``really good'' (see \cref{fig:boxplotSmileyometer}). The activities~1 and 3 are rated best in the Again Again table with \againOneAllMean on average and activities~5 and 7 are rated worst with \againFiveAllMean on average. In the Smileyometer, activity~4 is rated best with \likeFourAllMean on average and activity~7 is rated worst with~\likeSevenAllMean. Even though there is no significant  difference, the individual bad ratings of activity~7 could be caused by different factors:
Activity~7 was the last activity of each workshop, which is why there often was insufficient time to transition properly from the pen-and-paper mode to the digital mode and the digital mode itself. This might have overwhelmed some children, and it might be interesting to research good ways of transitioning and how this might affect the experienced fun when programming digitally.
Another explanation might be that some students might not have had sufficient time to draw their own path in the pen-and-paper mode (because they did not complete other activities early enough to do this additional task). Nevertheless the ratings for activity~7 are still good (see \cref{fig:boxplotAgain} and \cref{fig:boxplotSmileyometer}) and the slight differences might just derive from the pen-and-paper mode being especially engaging.


\subsubsection{Relation between Problems and Fun}

\input{tables/problemsFun}

\cref{tab:problemsFun} lists the means of the fun ratings associated with having received feedback on specific problems, and the correlation between fun and number of problems of a type. As is common the case with survey
questions for this target age group, the ratings are generally very positive
(see \cref{fig:boxplotAgain} and \cref{fig:boxplotSmileyometer}). \Cref{tab:problemsFun} shows that having experienced a problem and
thus having received feedback can however lead to worse fun ratings (indicated by $r_{pb}$ values above 0) but also to better fun ratings (indicated by $r_{pb}$ values below 0). We therefore take a closer qualitative look at the individual means and
relate them to the groups' problems.

\looseness=-1

Receiving feedback on rather technical and organisational issues (``P8: Other'') did not affect the experienced fun negatively (Again Again table: $p = \correlationAgainProblemOtherPValue$, $r_{pb} = \correlationAgainProblemOther$ ; Smileyometer: $p = \correlationLikeProblemOtherPValue$, $r_{pb} = \correlationLikeProblemOther$). One explanation might be that, e.g., an empty battery does not harm self-efficacy as these problems are out of the students' control and rather easy to solve.
For all other problems, students might attribute the problem, and with that the corrective feedback, to their own abilities. This might lead to a reduced intrinsic motivation as the students might not feel competent and autonomous~\cite{ryan2000intrinsic}. 

Within the problems that might be attributed to the students' abilities, the urgency of the problem might explain most differences regarding how much the activities were liked. Many problems are less urgent such as Ozobot Evo specific problems (P3), or U-turns (P5), as they are rather concerned with one specific code or concept and the students might meanwhile perform another part of the activity.
However, if the task is not understood (P1) or the app cannot be used (P7), the
students are not able to proceed. When having experienced one of these problems, the students rated the likeability of the respective activity significantly worse (P1: $p = \correlationLikeProblemUnderstandingPValue$, $r_{pb} = \correlationLikeProblemUnderstanding$; P7: $p = \correlationLikeProblemAppPValue$, $r_{pb} = \correlationLikeProblemApp$).
As supporting students is a time-consuming
and common challenge in the computer science
classroom~\cite{yadav2016expanding,michaeli2019current}, students might have to
wait for supervisors or teachers. 
It seems that this does not strongly affect the students' intention to perform the activity again but it comparatively often negatively affects the rating if the students liked it (\cref{tab:problemsFun}). This might be because having to wait for corrective feedback to proceed might be frustrating~\cite{wiggins2012seven} and thus reduce the likeability of the activity. 
These observations match the general result that having received feedback led to significantly worse ratings with the Smileyometer than not having received feedback ($p = \correlationAllTasksLikePValue$, $r_{pb} = \correlationAllTasksLike$).
Our data of the Again Again table (\cref{tab:problemsFun}) however indicate that fun and learning do not have to be mutually exclusive:
Besides likeability (measured with the Smileyometer), the endurability of the engagement (measured with the Again Again table) is another characteristic of fun activities~\cite{read2008validating}.
Our results show that the students want to have further time to apply and practice the respective activity again which might indicate an interest in learning more about it.
This matches findings that having received corrective feedback helps to acquire further skills and knowledge~\cite{wisniewski2020power} and thus enables learning. 
As the workshop aims at several learning objectives (see \cref{sec:workshop}), giving corrective feedback is crucial. 
However, the way of how to give corrective feedback should be considered.


\subsubsection{Relation between Tutoring Component and Fun}
\input{tables/tutoringFun}

\cref{tab:tutoringFun} lists the means of the fun ratings associated with having received feedback via a specific tutoring component, and the correlation between fun and how much of the feedback type was received. It shows that having received tutoring does not necessarily lead to worse fun ratings. We again take a closer qualitative look at the individual means and relate them to the tutoring component.

\looseness=-1
The mean ratings of the Again Again table are not substantially affected by tutoring components (\cref{tab:tutoringFun}). For the Smileyometer, there are statistically significant differences: When students received direct instruction, the activity is liked significantly less ($p = \correlationLikeTutoringDirectInstructionPValue$, $r_{pb} = \correlationLikeTutoringDirectInstruction$). This matches results that direct instruction is less motivating than, e.g., more cooperative or project-based learning~\cite{hanze2007cooperative,carrabba2018impact}. However, direct instruction is proven to be very effective, which is why modern and less strict forms such as Explicit Direct Instruction in Programming suggested by Hermans and Smit might be an appropriate solution~\cite{hermans2018explicit}.

Hints, on the contrary, affected the Smileyometer ratings in a positive way ($p = \correlationLikeTutoringHintPValue$, $r_{pb} = \correlationLikeTutoringHint$). Giving hints involves the children to think and thus gives them a partial autonomy in their learning which might increase their intrinsic motivation~\cite{ryan2000intrinsic}. As is known from automated hints during programming, hints are especially effective when they include a self-explanation prompt~\cite{marwan2019evaluation}.

\vspace{1em}
\summary{RQ2}{Our results indicate that corrective feedback can lead to worse fun ratings if the feedback is urgently needed and might harm self-efficacy. Tutoring components such as direct instruction can lead to worse, and hints to better fun ratings. While the intention to perform an activity again tends to be only slightly reduced, the likeability of an activity can be significantly reduced by having received corrective feedback.}

\subsection{RQ 3: Effects of Gender}

\input{tables/funGender}

\input{tables/pvaluesProblems}

Since introducing programming at primary school level is one of the strategies
of addressing the issue of underrepresented groups, we are specifically
interested in understanding whether the problems encountered differ based on
gender-related group constellations. \Cref{tab:pvaluesProblems} shows whether the group constellation had an effect on the
number of a specific problem. 

There is only one statistically significant difference:
Mixed groups needed feedback significantly more often on the problem of moving or turning into a specific direction (P6) ($p = \problemDirectionGroupspValue$, $\eta^{2} = \problemDirectionGroupsEffectSize$).
Regarding the overall number of experienced problems, there is no statistically significant difference between all-female, all-male and mixed groups ($p = \amountInteractionsGroupspValue$, $\eta^{2} = \amountInteractionsGroupsEffectSize$).
Thus, the workshop tends to be equally challenging for all group constellations. Other studies found gender differences for more complex tasks, but basic tasks were also performed similarly well~\cite{sullivan2013, sullivan2016}. Our experiences match these results as the Ozobot Evo programming activities deal with introductory tasks and topics such as sequences.


Considering fun, on average over all activities girls and boys liked programming with
Ozobot Evo robots similarly with the Smileyometer mean of
\allActivitiesFemaleLikeMean for female and \allActivitiesMaleLikeMean for male
students (\cref{tab:funGender}). Only for activity~5, boys rated it significantly better with the Again Again table than girls ($p = \againFiveGenderpValue$, $r = \againFiveGenderEffectSize$). This activity includes more letting the Ozobot Evo drive and less actively drawing something (see \cref{fig:overview}) which might cause gender-specific differences.
For the group constellations there are no significant differences but very similar results for all-female groups (Again Again table: \allActivitiesTwoFemalesAgainMean, Smileyometer: \allActivitiesTwoFemalesLikeMean), all-male groups (Again Again table: \allActivitiesTwoMalesAgainMean, Smileyometer: \allActivitiesTwoMalesLikeMean) and mixed groups (Again Again table: \allActivitiesFemaleMaleAgainMean, Smileyometer: \allActivitiesFemaleMaleLikeMean).
The workshop tends to be very enjoyable for both girls and boys. That matches former results where the Ozobot robot raised the interest of young female learners~\cite{french2018using}.

\vspace{1em}
\summary{RQ3}{Programming the Ozobot Evo is suitable for encouraging girls as the course was evaluated positively regardless of gender. Our results indicate that mixed groups might need more feedback on directional movements.}

%% file: tables/heatmapProblemsActivities.tex
\begin{table}[t]
\centering
\vspace{-0.5em}
\caption{Problems encountered by the students regarding the individual activities.}
\vspace{-0.5em}
	\label{tab:heatmapProblemsActivities}
\begin{tabular}{rcccccccccc}
\toprule

& \multicolumn{1}{c}{P1} & \multicolumn{1}{c}{P2} & \multicolumn{1}{c}{P3} & \multicolumn{1}{c}{P4} & \multicolumn{1}{c}{P5} & \multicolumn{1}{c}{P6} & \multicolumn{1}{c}{P7} & \multicolumn{1}{c}{P8}\\

\toprule

Activity 1 & 4 & 0 & \cellcolor{gray!10}{5} & 0 & 0 & 0 & 0 & 0 \\

Activity 2 & \cellcolor{gray!10}{6} & 0 & \cellcolor{gray!10}{7} & 0 & 0 & 0 & 0 & 0 \\

Activity 3 & 4 & 2 & 1 & 1 & 0 & 0 & 0 & 2 \\

Activity 4 & 1 & 4 & 0 & 0 & 0 & 0 & 0 & 0 \\

Activity 5 & \cellcolor{gray!10}{6} & \cellcolor{gray!20}{10} & \cellcolor{gray!10}{8} & 4 & 2 & 3 & 0 & 2 \\

Activity 6 & \cellcolor{gray!20}{17} & \cellcolor{gray!40}{26} & \cellcolor{gray!10}{8} & \cellcolor{gray!30}{21} & \cellcolor{gray!40}{25} & \cellcolor{gray!10}{10} & 0 & 0 \\

Activity 7 & \cellcolor{gray!10}{9} & 0 & 2 & 0 & 0 & 3 & \cellcolor{gray!20}{15} & \cellcolor{gray!10}{6} \\
\bottomrule
\end{tabular}
\end{table}

%% file: tables/heatmapProblemsFeedback.tex
\begin{table}[t]
\centering
\vspace{-0.5em}
\caption{Tutoring feedback type given by the supervisors ordered by frequency and related to the problems experienced by the primary school children.}
\vspace{-0.5em}
	\label{tab:heatmapProblemsFeedback}
\begin{tabular}{lcccccccccc}
\toprule

& \multicolumn{1}{c}{P1} & \multicolumn{1}{c}{P2} & \multicolumn{1}{c}{P3} & \multicolumn{1}{c}{P4} & \multicolumn{1}{c}{P5} & \multicolumn{1}{c}{P6} & \multicolumn{1}{c}{P7} & \multicolumn{1}{c}{P8}\\

\toprule

T1: Explanations & \cellcolor{gray!40}{21} & \cellcolor{gray!10}{7} & \cellcolor{gray!20}{14} & \cellcolor{gray!30}{15} & \cellcolor{gray!20}{11} & \cellcolor{gray!20}{12} & 2 & 2 \\

T2: Stickers & 0 & \cellcolor{gray!50}{38} & 2 & \cellcolor{gray!20}{11} & \cellcolor{gray!10}{8} & \cellcolor{gray!10}{5} & 0 & 0 \\

T3: Direct Instruction & \cellcolor{gray!20}{10} & 0 & \cellcolor{gray!10}{8} & 1 & 1 & 1 & \cellcolor{gray!20}{10} & 2 \\

T4: Hints  & \cellcolor{gray!20}{10} & 2 & 3 & \cellcolor{gray!10}{7} & \cellcolor{gray!10}{7} & 1 & 2 & 0 \\

T5: Other & \cellcolor{gray!10}{6} & 0 & 4 & 1 & 3 & 0 & 1 & \cellcolor{gray!10}{6} \\

T6: Guiding questions & \cellcolor{gray!10}{9} & 1 & 1 & 0 & \cellcolor{gray!10}{5} & 2 & 0 & 0 \\

\bottomrule

\end{tabular}
\end{table}

%% file: figures/boxplotAgain.tex
\begin{figure}[tb]
	\centering
	\includegraphics[width=\columnwidth]{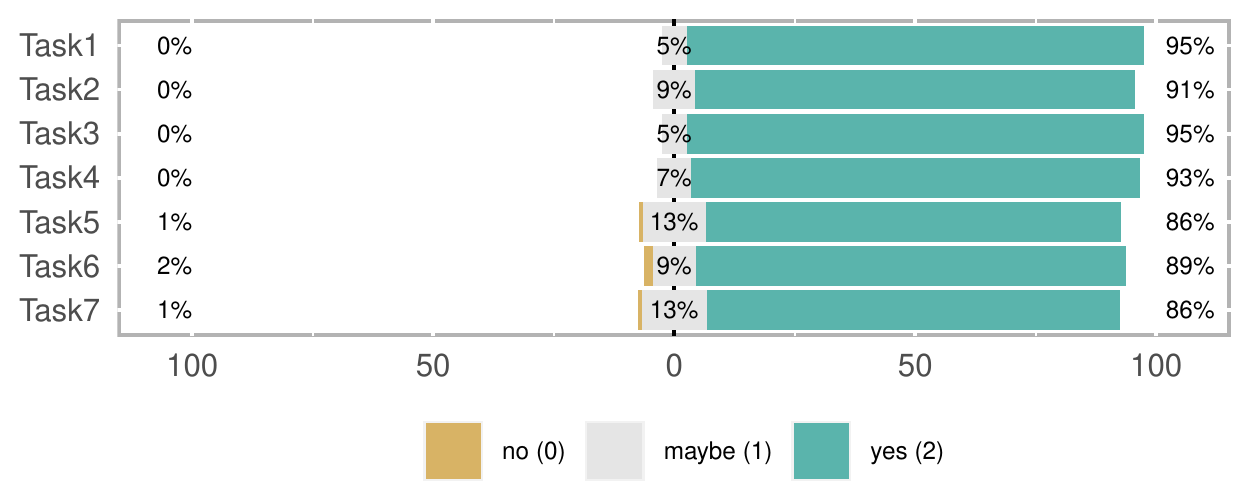}
	\caption{\label{fig:boxplotAgain}Values of the Again Again table.}
\end{figure}

%% file: figures/boxplotLike.tex
\begin{figure}[tb]
	\centering
	\includegraphics[width=\columnwidth]{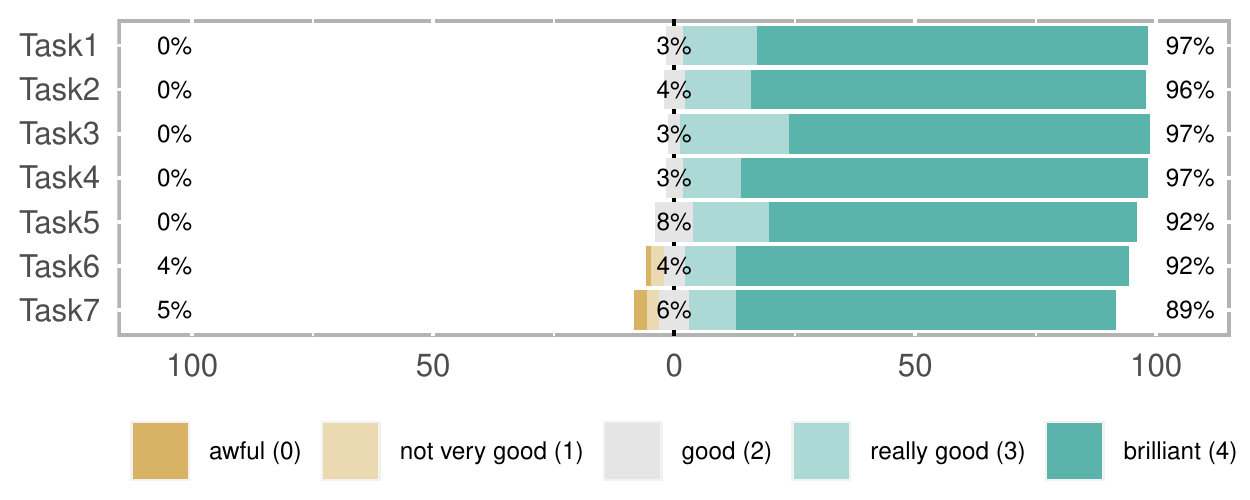}
	\caption{\label{fig:boxplotSmileyometer}Values of the Smileyometer.}
\end{figure}

%% file: tables/problemsFun.tex
\begin{table}[t]
\centering

\caption{Number of problems and associated fun ratings. \\
	\textmd{\small AT = Again Again Table, S = Smileyometer \\
	Coefficients associated with statistically significant $p$-values are bold.}}
	\label{tab:problemsFun}
	\resizebox{\columnwidth}{!}{

\begin{tabular}{lrrrrr}

\toprule

Problem & \#Total & AT & $r_{pb}$ & S & $r_{pb}$ \\

\midrule

P1: Task understanding & \frequencyProblemtypeUnderstandingofthetask & \meanAgainProblemtypeUnderstandingofthetask & \correlationAgainProblemUnderstanding & \meanLikeProblemtypeUnderstandingofthetask & \textbf{\correlationLikeProblemUnderstanding} \\

P2: Wrong colour code & \frequencyProblemtypeWrongcolourcode & \meanAgainProblemtypeWrongcolourcode & \correlationAgainProblemWrongCode & \meanLikeProblemtypeWrongcolourcode & \correlationLikeProblemWrongCode \\

P3: Ozobot Evo specific & \frequencyProblemtypeOzobotspezificproblems & \meanAgainProblemtypeOzobotspezificproblems & \correlationAgainProblemSpecific & \meanLikeProblemtypeOzobotspezificproblems & \textbf{\correlationLikeProblemSpecific} \\

P4: Reading direction & \frequencyProblemtypeReadingdirection & \meanAgainProblemtypeReadingdirection & \correlationAgainProblemReading & \meanLikeProblemtypeReadingdirection & \correlationLikeProblemReading \\

P5: U-turns & \frequencyProblemtypeUturn & \meanAgainProblemtypeUturn & \correlationAgainProblemUturn & \meanLikeProblemtypeUturn & \textbf{\correlationLikeProblemUturn} \\

P6: Direction & \frequencyProblemtypeDirection & \meanAgainProblemtypeDirection & \correlationAgainProblemDirection & \meanLikeProblemtypeDirection & \correlationLikeProblemDirection \\

P7: App & \frequencyProblemtypeApp & \meanAgainProblemtypeApp & \correlationAgainProblemApp & \meanLikeProblemtypeApp & \textbf{\correlationLikeProblemApp} \\

P8: Other & \frequencyProblemtypeOther & \meanAgainProblemtypeOther & \correlationAgainProblemOther & \meanLikeProblemtypeOther & \correlationLikeProblemOther \\

\bottomrule
\end{tabular}
}
\end{table}

%% file: tables/tutoringFun.tex
\begin{table}[t]
\centering
\caption{Number of tutoring type and associated fun ratings.	\textmd{\small AT = Again Again Table, S = Smileyometer \\
Coefficients associated with statistically significant $p$-values are bold.}}
	\label{tab:tutoringFun}
	\vspace{-0.5em}
	\resizebox{\columnwidth}{!}{

\begin{tabular}{lrrrrr}

\toprule

Tutoring type & \#Total & AT & $r_{pb}$ & S & $r_{pb}$ \\

\midrule

T1: Explanations & \frequencyTutoringcomponentExplanations & \meanAgainTutoringcomponentExplanations & \correlationAgainTutoringDirectInstruction & \meanLikeTutoringcomponentExplanations & \correlationLikeTutoringExplanations \\

T2: Stickers & \frequencyTutoringcomponentSticker & \meanAgainTutoringcomponentSticker & \correlationAgainTutoringSticker & \meanLikeTutoringcomponentSticker & \correlationLikeTutoringSticker \\

T3: Direct instruction & \frequencyTutoringcomponentDirectInstruction & \meanAgainTutoringcomponentDirectInstruction & \correlationAgainTutoringDirectInstruction & \meanLikeTutoringcomponentDirectInstruction & \textbf{\correlationLikeTutoringDirectInstruction} \\

T4: Hints & \frequencyTutoringcomponentHint & \meanAgainTutoringcomponentHint & \correlationAgainTutoringHint & \meanLikeTutoringcomponentHint & \textbf{\correlationLikeTutoringHint}  \\

T5: Other & \frequencyTutoringcomponentOther & \meanAgainTutoringcomponentOther & \correlationAgainTutoringOther & \meanLikeTutoringcomponentOther & \correlationLikeTutoringOther \\

T6: Guiding questions & \frequencyTutoringcomponentGuidingquestions & \meanAgainTutoringcomponentGuidingquestions & \correlationAgainTutoringGuiding & \meanLikeTutoringcomponentGuidingquestions & \correlationLikeTutoringGuiding \\

\bottomrule
\end{tabular}
}
\vspace*{-9pt}
\end{table}

%% file: tables/funGender.tex
\begin{table}[t]
\centering
\caption{Means of fun for each activity for girls and boys. \\
	\textmd{\small AT = Again Again Table, S = Smileyometer \\ m = male students, f = female students}}
	\vspace{-0.5em}
	\label{tab:funGender}
	\resizebox{\columnwidth}{!}{

\begin{tabular}{rrrrrrrrrrrrrrr}

\toprule
& \multicolumn{2}{c}{A1} & \multicolumn{2}{c}{A2} & \multicolumn{2}{c}{A3} & \multicolumn{2}{c}{A4} & \multicolumn{2}{c}{A5} & \multicolumn{2}{c}{A6} & \multicolumn{2}{c}{A7} \\
& AT & S & AT & S & AT & S & AT & S & AT & S & AT & S & AT & S  \\

\midrule

m & \againOneMaleMean & \likeOneMaleMean & \againTwoMaleMean & \likeTwoMaleMean & \againThreeMaleMean & \likeThreeMaleMean & \againFourMaleMean & \likeFourMaleMean & \againFiveMaleMean & \likeFiveMaleMean & \againSixMaleMean & \likeSixMaleMean & \againSevenMaleMean & \likeSevenMaleMean \\

f & \againOneFemaleMean & \likeOneFemaleMean & \againTwoFemaleMean & \likeTwoFemaleMean & \againThreeFemaleMean & \likeThreeFemaleMean & \againFourFemaleMean & \likeFourFemaleMean & \againFiveFemaleMean & \likeFiveFemaleMean & \againSixFemaleMean & \likeSixFemaleMean & \againSevenFemaleMean & \likeSevenFemaleMean \\

\bottomrule
\end{tabular}
}
\end{table}

%% file: tables/pvaluesProblems.tex
\begin{table}[t]
\centering
	\caption{P-values, effect sizes and means of encountered problems regarding group constellations. \\
	\textmd{m2 = all-male groups, f2 = all-female groups, fm = mixed groups}}
	\vspace{-0.5em}
	\label{tab:pvaluesProblems}
	\resizebox{\columnwidth}{!}{
\begin{tabular}{rrrrrrrrrr}

\toprule
& P1 & P2 & P3 & P4 & P5 & P6 & P7 & P8 & P1 to P8 \\

\midrule
$p$ & \problemUnderstandingTaskGroupspValue & \problemWrongCodeGroupspValue & \problemOzobotSpecificGroupspValue & \problemReadingDirectionGroupspValue & \problemUTurnGroupspValue & \problemDirectionGroupspValue & \problemAppGroupspValue & \problemOtherGroupspValue & \amountInteractionsGroupspValue \\

$\eta^{2} $ & \problemUnderstandingTaskGroupsEffectSize & \problemWrongCodeGroupsEffectSize & \problemOzobotSpecificGroupsEffectSize & \problemReadingDirectionGroupsEffectSize & \problemUTurnGroupsEffectSize & \problemDirectionGroupsEffectSize & \problemAppGroupsEffectSize & \problemOtherGroupsEffectSize & \amountInteractionsGroupsEffectSize \\

m2 & \problemUnderstandingTaskGroupsTwoMales & \problemWrongCodeGroupsTwoMales & \problemOzobotSpecificGroupsTwoMales & \problemReadingDirectionGroupsTwoMales & \problemUTurnGroupsTwoMales & \problemDirectionGroupsTwoMales & \problemAppGroupsTwoMales & \problemOtherGroupsTwoMales & \amountInteractionsGroupsTwoMalesMean \\

f2 & \problemUnderstandingTaskGroupsTwoFemales & \problemWrongCodeGroupsTwoFemales & \problemOzobotSpecificGroupsTwoFemales & \problemReadingDirectionGroupsTwoFemales & \problemUTurnGroupsTwoFemales & \problemDirectionGroupsTwoFemales & \problemAppGroupsTwoFemales & \problemOtherGroupsTwoFemales & \amountInteractionsGroupsTwoFemalesMean \\

fm & \problemUnderstandingTaskGroupsMixed & \problemWrongCodeGroupsMixed & \problemOzobotSpecificGroupsMixed & \problemReadingDirectionGroupsMixed & \problemUTurnGroupsMixed & \problemDirectionGroupsMixed & \problemAppGroupsMixed & \problemOtherGroupsMixed & \amountInteractionsGroupsMixedMean \\

\bottomrule
\end{tabular}
}
\end{table}

%% file: content/discussion.tex
\section{Discussion}
\label{sec:discussion}

We found several common problems (RQ 1) that can lead to worse fun ratings when the feedback on them is urgently needed or when the feedback is given via direct instruction (RQ 2).
As the fun ratings are generally very high, this course fulfils its goal of fostering children's initial interest in programming, both for girls and boys (RQ 3).
Different categories of motivation appear to play a role while children are acting, learning and achieving goals. One category deals with explaining children's engagement and investment. Since enjoyment is a well established reason in this context (see Eccles' Expectancy-Value Theory)~\cite{wigfield2006development}, we investigate the development of fun during our intervention. Further research is needed to examine the relations between positive emotions and learning when giving corrective feedback to students working with educational robots.

\subsection{Addressing Common Problems in Advance}
%
In RQ~1, we explored common problems so that teachers can build knowledge on them and prepare auxiliary material in advance to counteract the challenge of supporting students~\cite{sentance2017computing,yadav2016expanding,greifenstein2021challenging}. 
Some problems could be addressed in class and we discuss if this applies to other programming activities, too.

To reduce the requests for feedback on the understanding of the task, a presentation or worksheet with task constraints might support students to remember and understand it. Parental support can also redirect to the task, e.g., when programming with the KIBO robotics kit~\cite{relkin2020}, and could be realised in the form of a visitor's day~\cite{greifenstein2021challenging}, while guiding questions on the task were rather overwhelming when programming Thymio robots~\cite{chevalier2022role}.

Introductory tasks often deal with robotic movements, e.g., for Lego Mindstorms, the Bee-Bot or the Thymio robot~\cite{jin2016teaching,beraza2010soft,riedo2013thymio}.
%
%
To understand the path of the robot, the students can be encouraged to put themselves into the view of the robot, walking the path step by step or rotating the map with the path.
Moreover, the start and end point could be emphasised verbally and/or highlighted in colour for the students to remember. 
For robots that include ``jumps'' in their features, such as the Ozobot Evo, one can make clear in advance---if the robots do not really jump in the air---that they do not have elastic springs. The actuators are often discussed as an initial activity of a robotic workshop~\cite{korber2021experience,riedo2013thymio}.

While the colour codes are pre-programmed for the Ozobot Evo, tasks for other robots also often deal with coloured lines, e.g., line-following or reacting to colours~\cite{riedo2013thymio,su2010intelligent,jin2016teaching}.
To provide an initial hands-on experience, some colour codes could be programmed by the educator for other robots, too.
This would allow sequences or loops to be easily visualised by the robot's path without immediately opening the black box. Moreover, the difference between static code (= path with colour codes) and running code dynamically (= robot moving along the path) could be addressed explicitly as it can be shown in a comprehensible way.
After that, for the transitioning from a non-digital to the digital programming mode, one could combine both modes by an animation: The colour codes could be replaced by commands and put in the form of a script. 

\subsection{Giving Encouraging Feedback}

Renninger~\cite{renninger2009interest} proposed a phase model of interest development, that starts with phases describing situational interest. According to this model, positive feelings are helpful to maintain situational interest. This matches the aim of promoting interest when programming in the primary school classroom~\cite{greifenstein2021challenging}.
%
While the Ozobot Evo and the workshop setting generally had positive effects on the children's fun, we still found differences depending on the specific problem and tutoring component given to the children in RQ~2.

To raise situational interest, feedback should ensure that the children are granted to solve their problem autonomously to some extent. We found that hints should be preferred over direct instruction. 
As the basic human needs of autonomy, relatedness, and competence are crucial for building intrinsic motivation~\cite{ryan2000intrinsic}, we  focussed on hints that empowered children to solve the problem without direct instruction of the teacher.
In our setting this is done in a co-constructive process enabling children to experience success when solving problems. 
Crucial for this process are hints that are reasonably balanced between being too unspecific and revealing the solution directly.
Timing is another critical factor that needs to be considered:
Feedback should on the one hand be given timely~\cite{wiggins2012seven}, but on the other hand slightly delayed so that the students are not relying solely on feedback~\cite{chevalier2022role}.
Finally we suggest that all information which cannot at all be figured out independently and/or is not crucial for the specific learning process should be given in front of the whole class rather than being part of an individual feedback. 



\subsection{Preparing Material for Individual Students}

In RQ~3, we found that gender has no major effect on problems or fun. Still, other factors of heterogeneity might cause different kind of problems such as some groups needing support in the earlier activities while most groups didn't (see \cref{tab:heatmapProblemsActivities}).
To reach all students, experienced teachers differentiate their learning material~\cite{greifenstein2021challenging} and we discuss how this applies to robot programming activities.

Not all specific problems should be addressed in detail in front of the class as students should first have the chance of solving a problem on their own. 
To support struggling students, auxiliary material can be given to them. This might save time during lessons, but preparing it of course takes time beforehand. 
When programming a robot requires using an app, general features such as connecting robots or deleting code should be explained. Additionally, explanations of further potential issues should be outsourced. One idea is to use edited screen captures addressing subproblems.

When struggling to find the correct (colour) code, hints or guiding questions could be printed out. There are somewhat ambiguous results regarding these tutoring feedback types: While hints when programming might be useful in an educational context~\cite{greifenstein2021effects,obermuller2021guiding}, effects of guiding questions might depend on their meta-cognitive complexity and the learners' self-direction~\cite{chevalier2022role}.
When wrong (colour) codes have been used, the code has to be tested and debugged iteratively. For the Ozobot Evo robots, stickers are a common way to handle colour code mistakes~\cite{fojtik2017ozobot} (\cref{tab:heatmapProblemsFeedback}). This refers to, e.g., deleting a block in a block-based programming environment.
Generally, it might be useful to have a dedicated desk where auxiliary material, e.g., stickers or explanatory sheets, are provided.

%% file: content/conclusions.tex
\section{Conclusions}
\label{sec:conclusions}

\looseness=-1
In order to explore what are the problems that primary school children encounter while working with Ozobot Evo robots, which corrective feedback can be used to address these problems, and to what extent this influences the fun, we conducted \numWorkshops workshops with \numParticipantsThirdGrade third and \numParticipantsFourthGrade fourth graders with the Ozobot Evo robot.
We observed that most of the problems were caused by task constraints and wrong (colour) codes. 
We found that the corrective feedback on the problems partially
influenced the fun---depending on the type of the problem and the tutoring component: We therefore suggest to give prepared hints rather than using direct instruction.

Overall, the course was evaluated very positively by the students, regardless of gender.
We conclude that an introductory programming course with Ozobot Evo robots can be equally enjoyable for girls and boys.
As a next step, we plan to evaluate auxiliary material for common problems when programming the Ozobot Evo and to study the effects of self-selected feedback on the students' self-efficacy, fun and learning when building and programming another robot.
Moreover, we suggest to offer more advanced subsequent courses---such as parent-child courses---in the future in order to promote a more well-developed individual interest.